*Article.*

# Tetrel Bonding in Anion Recognition: A First-Principles Investigation

**Pradeep R. Varadwaj**[1,2][*]

1. Department of Chemical System Engineering, School of Engineering, The University of Tokyo 7-3-1, Tokyo 113-8656, Japan
2. Molecular Sciences Institute, School of Chemistry, University of the Witwatersrand, Johannesburg 2050, South Africa
* Correspondence: pradeep@t.okayama-u.ac.jp

**Abstract:** Twenty-five molecule-anion complex systems [$I_4Tt$···$X^-$] (Tt = C, Si, Ge, Sn and Pb; X = F, Cl, Br, I, At) were examined using density functional theory (ωB97X-D) and *ab initio* (MP2 and CCSD) methods to demonstrate the ability of the tetrel atoms in molecular entities, $I_4Tt$, to recognize the halide anions when in close proximity. The tetrel bond strength for the [$I_4C$···$X$]$^-$ series, and [$I_4Tt$···$X$]$^-$ (Tt = Si, Sn; X = I, At), was weak-to-moderate, whereas that in the remaining 16 complexes was dative tetrel bond type with very large interaction energies and short Tt···X close contact distances. The basis set superposition error corrected interaction energies calculated with the highest-level theory applied, [CCSD(T)/def2-TZVPPD], ranged from –3.0 to –112.2 kcal mol$^{-1}$. The significant variation in interaction energies was realized as a result of different levels of tetrel bonding environment between the interacting partners at the equilibrium geometries of the complex systems. Although the ωB97X-D computed intermolecular geometries and interaction energies of [$I_4Tt$···$X^-$] were close to those predicted by the highest level of theory, the MP2 results were shown to be misleading for some of the molecule-anion complex systems investigated. To provide insight into the nature of the intermolecular chemical bonding environment in the 25 molecule-anion complexes investigated, we discussed the charge density-based topological and isosurface features that emanated from the application of quantum theory of atoms in molecules and independent gradient model approaches, respectively.

**Keywords:** Tetrel bond; non-covalent interactions; weak-to-strong interaction energy; dative bond formation; chemical bonding; anion recognition; MESP analysis; charge-density analysis; first-principles calculations

## 1. Introduction

Ion-molecule interactions are fascinating in chemistry [1–4], biology [5] and materials science [6–8]. These interactions are ubiquitous in many chemical systems in solid, liquid, and gas phases, and play an important role in sensing, extraction, transport, assembly and catalysis [6]. They appear between an anion at the molecular (or atomic) level and a neutral molecule, or between a cation and a neutral molecule. The Cambridge Structural Database (CSD) [9] has catalogued many such chemical systems in the crystalline phase [10,11].

Figure 1a-f, for example, provides experimental evidence of molecule-anion systems in the crystalline phase. In them, the halide anion (Cl$^-$, or Br$^-$, or I$^-$) attracts the electrophiles on bonded Si atoms in the neutral molecules. Neither anion is precisely at the centroid of the neutral molecule. When the anion (Cl$^-$) is entrapped inside the Si$_{20}$ cage of a fully, or partially, chlorinated, icosasilane molecular entity (Figures 1e and 1f), its position is also off-center, so that it could maximize its attraction with all the 20 Si atoms of the Si$_{20}$ cage to stabilize the (Cl–)Si···Cl and/or (H–)Si···Cl close contacts.

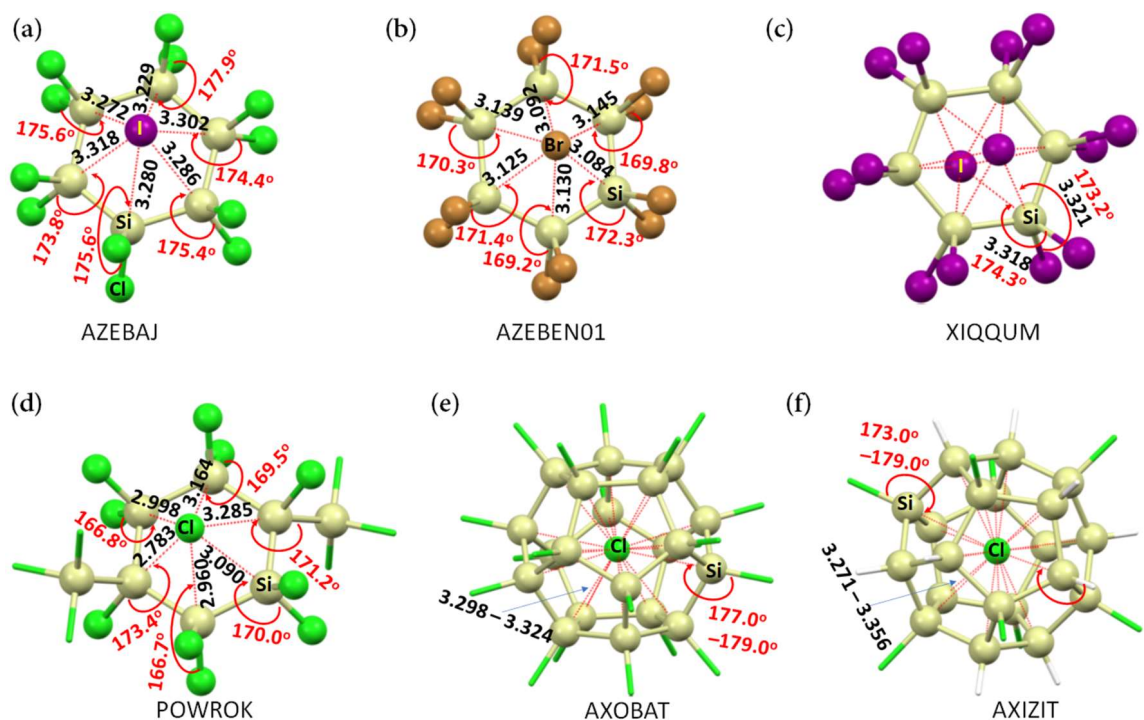

**Figure 1.** Illustration of molecule-anion interactions in the anionic-part of some selected chemical systems catalogued in the CSD[9]. The crystals include: (a) bis(tetrabutylammonium) dodecachlorohexasilinane bis(iodide) dichloromethane solvate [2($C_{16}H_{36}N^+$),$Cl_{12}Si_6$,$CH_2Cl_2$,2($I^-$)][12]; (b) bis(tetrabutylammonium) bis(bromide) dodecabromohexasilinane [2($C_{16}H_{36}N^+$),$Br_{12}Si_6$,2($Br^-$)][13]; (c) bis(tetrabutylphosphanium) dodecaiodohexasilinane bis(iodide) [2($C_{16}H_{36}P^+$),$I_{12}Si_6$,2($I^-$)][13]; (d) bis(tetraphenylphosphonium) 1,1,2,2,3,4,4,5,5,6-decachloro-3,6-bis(trichlorosilyl)hexasilinane bis(chloride) [2($C_{24}H_{20}P^+$),$Cl_{16}Si_8$,2($Cl^-$)][14]; (e) triphenyl-N-(triphenylphosphanylidene)phosphaniminium chloride 1,2,3,4,5,6,7,8,9,10,11,12,13,14,15,16,17,18,19,20-icosachloroundecacyclo[9.9.0.0$^{2,9}$.0$^{3,7}$.0$^{4,20}$.0$^{5,18}$.0$^{6,16}$.0$^{8,15}$.0$^{10,14}$.0$^{12,19}$.0$^{13,17}$]icosasilane chloroform solvate [$C_{36}H_{30}NP_2^+$,$Cl_{20}Si_{20}$,2($CHCl_3$),$Cl^-$][15]; (f) bis(triphenyl-N-(triphenylphosphanylidene)phosphaniminium) chloride 4-methylbenzene-1-sulfonate 1,3,5,8,10,13,16,19-octachloroundecacyclo[9.9.0.0$^{2,9}$.0$^{3,7}$.0$^{4,20}$.0$^{5,18}$.0$^{6,16}$.0$^{8,15}$.0$^{10,14}$.0$^{12,19}$.0$^{13,17}$]icosasilane [2($C_{36}H_{30}NP_2^+$),$H_{12}Cl_8Si_{20}$,$C_7H_7O_3S^-$,$Cl^-$][15]. The CSD reference codes are depicted in uppercase letters. Selected bond distances and bond angles associated with the (H–)Si···Cl and/or (X–)Si···X (X = Cl Br, I) are in Å and degree, respectively. ~~The anionic moieties in crystals (a)–(f) are omitted for clarity.~~ Atom labeling is shown for selected atoms.

The fundamental phenomena that drive isolated molecules to self-assemble with anions play an significant role in the processes of anion recognition and anion transport, among other [16–18]. One such phenomenon is the so-called intermolecular interactions, which are inherently noncovalent [18–20].

This study has theoretically examined 25 molecule-anion systems, including their intermolecular geometries, energies, and topological charge-density properties. The molecular entities were the heaviest members, $TtI_4$, of the tetrel tetrahalide family ($TtX_4$), where Tt stands for the elements in Group 14 of the periodic table and X represents the halide derivative (Tt = C, Si, Ge, Sn, Pb; X = F, Cl, Br, I, At). The anions considered were the halide derivative, $X^-$. It is worth mentioning that the theoretical chemistry of 1:1 complexes formed of lighter members of the $TtX_4$ (Tt = Si, Ge, Sn) family with the first three halide anions was recently reported [21–24]. However, the molecule–anion systems considered in this study have never been explored, probably because they were computationally intensive and require a large basis set due to the diffusivity of the heavy atoms involved.

The main purpose of this study is to theoretically clarify the following questions. i) How strong is the electrophilic region on the electrostatic surface of the Tt atom in molecular $TtI_4$? ii) Can electrophile on Tt be active enough to recognize the halide anions when in close proximity? iii) If so, what would be the strength of the interaction between the them? iv) Should we expect a dependence between descriptors of intermolecular interactions, such the Tt···X intermolecular distance and the interaction energy? v) Should the resulting intermolecular interactions between molecular entities responsible for the equilibrium geometries of the [I$_4$Tt···X$^-$] complexes be called ordinary tetrel bonds [23,25–27] or coordinative tetrel bonds [18,28]? A tetrel bond occurs in chemical systems when there is evidence of a net attractive interaction between an electrophilic region associated with a bound tetrel atom in a molecular entity and a nucleophilic region in another, or the same, molecular entity [29]. The chemical origin of a tetrel bond can be intermolecular, or intramolecular, and is formed by the elements of Group 14 in proximity to a nucleophile.

A number of computational approaches were employed to shed light on the set of questions posed above. The Molecular Electrostatic Surface Potential (MESP) [30–33] analysis was performed to determine the electrophilic and nucleophilic characters [34–39] of specific regions on the surface each molecular entity, TtI$_4$. Quantum Theory of Atoms in Molecules (QTAIM) [40–43] and Independent Gradient Model (IGM) [44,45]-based charge density analyses were performed to characterize the shared- and closed-shell characters of intra- and inter-molecular interactions [35,46] responsible for [I$_4$Tt···X$^-$]. Density functional theory (ωB97X-D [47]) and *ab initio* calculations (MP2 and CCSD) were performed to obtain geometries and electronic properties; MP2 and CCSD refer to the second-order Møller–Plesset theory [48–50] and Coupled Cluster theory with Singles and Doubles excitations [51,52], respectively.

## 2. Computational Methods

The Gaussian 16[53] calculator was used for the geometric relaxation of the 25 molecule-anion systems; the MP2, CCSD, and ωB97X-D approaches were employed. Two different basis sets were used, including def2-TZVPPD and def2-QZVPPD, obtained from the Basis Set Exchange library [54,55]. It is worth mentioning that we initially planned to use the former pseudopotential, together with the MP2 and ωB97X-D methods, in our calculations, and the MP2 method was chosen based on the results of many previous studies [18,22,56,57]. However, a large discrepancy between the trend in the MP2 and ωB97X-D interaction energies was found for a certain molecule-anion systems; this was due to the different nature of the tetrel bonding environment and unusually large basis set superposition error (BSSE) encountered with the post Hartree–Fock method. Therefore, we used a relatively large pseudopotential of quadrupole zeta-valence quality (def2-QZVPPD) to reexamine the MP2 geometries and energies of [I$_4$Tt···X$^-$]. Computationally expensive CCSD and CCSD(T) methods, in conjunction with the def2-TZVPPD basis set, were also employed to demonstrate the accuracy of MP2 and ωB97X-D geometries and energies of [I$_4$Tt···X$^-$]. Standard non-relativistic calculations were performed without considering the effect of spin-orbit coupling for heavy atoms such as Pb, following a previous study [17]. Default cutoff criteria for force and displacement for convergence of geometry and frequency calculations were considered. The eigenvalues associated with the normal mode vibrational frequencies of the isolated and complexed systems were all positive, thus the monomer and complex geometries reported are local minima.

To discuss the electrophilicity of the tetrel atom in TtI$_4$, the MESP analysis was performed on each of the five isolated monomer geometries. The MESP calculation has utilized wavefunctions generated by [ωB97X-D/def2-QZVPPD]. An isoelectron density envelope of 0.001 a.u. was used on which to compute the potential, even though the use of higher isoelectron density envelopes was suggested in other studies for chemical systems containing low-polarizable atomic basins [37,58,59]. We have done so as to obtain the sign and magnitude of local most maxima and minima of potential ($V_{S,max}$ and $V_{S,min}$, respectively)[37,46,60–62] on the electrostatic surfaces of molecular TtI$_4$. Gaussian 16 [53], Multiwfn [63], and VMD[64] codes were used.

Based on the basic concept of MESP [37,58,65,66], if the sign of either $V_{S,max}$ or $V_{S,min}$ on a specific region of the molecular surface is positive, then the region can be characterized to be electrophilic; if the sign of either $V_{S,max}$ or $V_{S,min}$ on a specific region of the molecular surface is negative, the region is characterized to be nucleophilic. It is often (but not always!) observed that the sign of $V_{S,max}$ is positive on the surface of an atom Tt opposite of the R–Tt covalent or coordinating bond; where R is the remainder part of the molecular entity. It occurs when R has a relatively stronger electron-withdrawing capacity than Tt, thus leaving a region of electron density deficiency on Tt on the opposite side of the R–Tt covalent or coordination bond. This electron density deficient region on Tt along the outer extension of the axial direction is referred to as a "σ-hole" [37,38,58,65,67,68]. It should be kept in mind that a σ-hole can be either positive or negative depending on whether $V_{S,max}$ is positive or negative. The coulombic attraction of an electrophilic σ-hole on the bonded Tt atom in R–Tt and a nucleophile on the same or a different molecular entity is referred to as a σ-hole-centered tetrel bond, or simply σ-hole a interaction [4,23,26,29].

The uncorrected and BSSE corrected interaction energies ($E_{int}$ and $E_{int}(BSSE)$, respectively) of each molecule-anion system were determined using Eqns. 1 and 2. In Eqn. 1, $E_T(complex)$, $E_T(iso_1)$, and $E_T(iso_2)$ represent the total electronic energies of the molecule-anion complex, isolated molecule, isolated anion, respectively; in Eqn. 2, $E(BSSE)$ is the error in the total electronic energy due to basis set superposition, obtained using the counterpoise procedure of Boys and Bernardi [69]. The geometry of the isolated molecule in the fully relaxed geometry of molecule-anion complex was used to obtain $E_T(iso_1)$.

$$E_{int} = E_T(complex) - E_T(iso_1) - E_T(iso_2) \quad (1)$$

$$E_{int}(BSSE) = E_{int} + E(BSSE) \quad (2)$$

QTAIM [40–43] calculations were performed for 25 molecule-anion systems using [ωB97X-D/def2-QZVPPD] geometries. Five bond descriptors were investigated, including the charge density $\rho_b$, the Laplacian of charge density ($\nabla^2\rho_b$), the gradient kinetic energy density ($G_b$), the potential energy density ($V_b$), and the total energy density $H_b$ ($H_b = V_b + G_b$) at bond critical points. The AIMAll code was used [70].

IGM[71,72]-based calculations were performed at the same theoretical level as QTAIM, and its implications have been actively discussed in various research papers [35,37,46]. The method was originally developed to use promolecular densities to explore the non-covalent chemistry of inter- and intra-molecular interactions in chemical and biological systems [71]. However, using actual densities [72] calculated at the [ωB97X-D/def2-QZVPPD] level, we show that the IGM-$\delta g^{inter}$ based isosurfaces between interacting atomic basins in [I₄Tt···X⁻] are consistent with the topological charge density-based features emanated using QTAIM. Both Multiwfn [63,73] and VMD[64] codes were used.

Delocalization index, δ, is a two-electron property, which is a measure of the number of electron pairs that are being shared between quantum atoms $\Omega_A$ and $\Omega_B$ [74,75]. It has also been interpreted as a measure of bond order [76], which is formally defined as the half the difference between the number of bonding and anti-bonding electrons [77]. We calculated δ within the framework of QTAIM to provide insight into the nature of tetrel bonds in [I₄Tt···X⁻]. The AIMAll code was used [70].

## 3. Results

### 3.1. The monomer properties

Molecular crystals of TtI₄ (Tt = C, Si, Ge, Sn) are known since the last century, yet there is no crystallographic evidence of molecular PbI₄. The crystal structure of the former four species can be retrieved from the Inorganic Chemistry Structure Database (ICSD) [78–82]. Table 1 lists the experimental bond distances $r$(Tt–I) and bond angles ∠I–Tt–I of molecular TtI₄, which are compared with those calculated with [MP2/def2-QZVPPD] and [ωB97X-D/def2-QZVPPD]. The best agreement between experiment and theory is observed with [ωB97X-D/def2-QZVPPD], and MP2 shows a tendency of un-derestimating $r$(Tt–I). A very similar trend was obtained with these methods in conjunction with the def2-QZVPPD basis set.

Table 1: Comparison of computed Tt–I bond distances $r$ (Å) and I–Tt–I bond angles ∠ (degree) of TtI$_4$ (Tt = C, Si, Ge, Sn and Pb) with those feasible in their corresponding crystals retrieved from the Inorganic Crystal Structure Database (ICSD).[a]

| Monomer | Property | Expt.[a] | [MP2/def2-QZVPPD] | [ωB97X-D/def2-QZVPPD] |
|---|---|---|---|---|
| **CI$_4$** | $r$(C–I) | 2.154 | 2.131 | 2.254 |
| | ∠I–C–I | 109.47 | 109.47 | 109.47 |
| **SiI$_4$** | $r$(Si–I) | 2.434 | 2.403 | 2.434 |
| | ∠I–Si–I | 109.47 | 109.47 | 109.47 |
| **GeI$_4$** | $r$(Ge–I) | 2.574 | 2.463 | 2.518 |
| | ∠I–Ge–I | 109.47 | 109.47 | 109.47 |
| **SnI$_4$** | $r$(Sn–I) | 2.650 | 2.463 | 2.518 |
| | ∠I–Sn–I | 109.47 | 109.47 | 109.47 |
| **PbI$_4$** | $r$(Pb–I) | --- | 2.705 | 2.749 |
| | ∠I–Pb–I | --- | 109.47 | 109.47 |

[a] CI$_4$ (ICSD ref: 30789); SiI$_4$ (ICSD ref: 22100); GeI$_4$ (ICSD ref: 22399); and SnI$_4$ (ICSD ref: 18010).

Figure 2a-e (Top) shows the molecular graph of isolated TtI$_4$. From these, it may be seen that the atomic basins are connected to each other in each isolated monomer

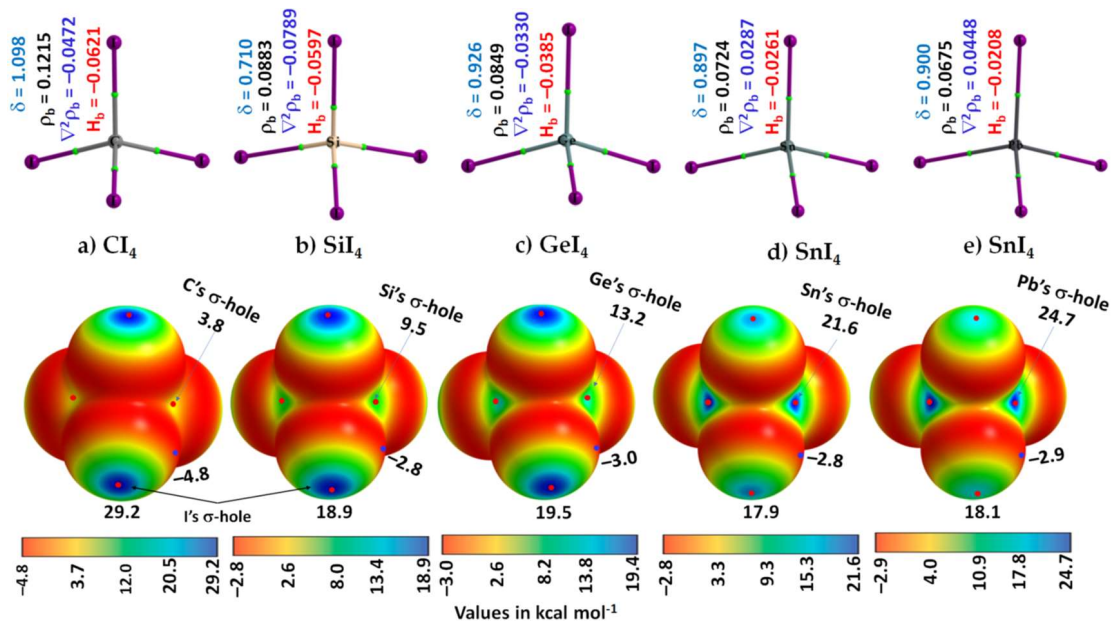

**Figure 2.** (Top) [ωB97X-D/def2-QZVPPD] level QTAIM-based molecular graphs of isolated TtI$_4$ (Tt = C, Si, Ge, Sn and Pb), showing bond paths (solid lines in atom color) and bond critical points (tiny spheres in green) between bonded atomic basins (large spheres). The charge density($\rho_b$), the Laplacian of the charge density ($\nabla^2\rho_b$), the total energy density (H$_b$) and the delocalization index (δ) values are shown in black, blue, red and faint-blue fonts (in a.u.), respectively. (Bottom) The 0.001 a.u. (electrons bohr$^{-3}$) isoelectron density mapped potential on the electrostatic surfaces of the corresponding monomers, including a) CI$_4$; b) SiI$_4$; c) GeI$_4$; d) SnI$_4$; and e) PbI$_4$. The strength of Tt's and I's σ-holes is shown in each case; filled tiny blue and red circles represent $V_{S,min}$ and $V_{S,max}$, respectively.

through bond paths (solid lines between atomic basins) that pass-through bond critical points (tiny green spheres), thus recovering the expected tetrahedral T$_d$ shape of TtI$_4$.

The charge density $\rho_b$ is larger at the C–I bcps in CI$_4$ than at the Pb–I bcps in PbI$_4$. It follows the trend across the series: $\rho_b$ (C–I) > $\rho_b$ (Si–I) > $\rho_b$ (Ge–I) > $\rho_b$ (Sn–I) > $\rho_b$ (Pb–I). The trend signifies that the charge concentration is predominant at the C–I bcps in CI$_4$ relative to that at the Pb–I bcps in PbI$_4$.

From the sign and magnitude of H$_b$, Figure 2a-e (Top), it may be seen that H$_b$ is stabilizing (H$_b$ < 0) at the Tt–I bcps, which is due to the potential energy density that dominates over the gradient kinetic energy density at the bcp. H$_b$ is increasingly more positive at the Tt–I bcps across the series from CI$_4$ through SiI$_4$ to GeI$_4$ to SnI$_4$ to PbI$_4$. This is consistent with the character of Tt–I bonds in TtI$_4$, in which, it becomes less covalent than ionic passing from CI$_4$ through SiI$_4$ to PbI$_4$. That is, the covalency of Tt–I bond follows this order: C–I > Si–I > Ge–I > Sn–I > Pb–I. Furthermore, the sign of $\nabla^2\rho_b$ at Tt–I bcps is also negative for all monomers except for the Tt–I (Tt = Sn, Pb) bcps, giving further indication that the Tt–I bonds in CI$_4$, SiI$_4$ and GeI$_4$ are relatively more covalent than those in SnI$_4$ and PbI$_4$. Typically, $\nabla^2\rho_b$ < 0 and H$_b$ < 0 represent covalent (shared-type) interactions; $\nabla^2\rho_b$ > 0 and H$_b$ > 0 represent ionic (closed-shell) interactions; and $\nabla^2\rho_b$ > 0 and H$_b$ < 0 represents mixed (ionic and covalent) interactions [83–88].

The delocalization indices, $\delta$, for atom-atom pairs involving Tt and I in TtI$_4$ ranged from 0.710 to 1.098, suggesting that they are bound to each other by a σ-type covalent (or coordinate) bond.

From the MESP graph, Figure 2a-e (Bottom), we observed that there are four σ-holes on each tetrel atom in TtI$_4$; they are appearing along the outer extensions of the four I–Tt covalent/coordinate bonds. They are equivalent for a given tetrel derivative in TtI$_4$ (two shown in each case). The strength of the σ-hole is quantified by the local most maximum of potential, $V_{S,max}$. It is varying from 3.8 kcal mol$^{-1}$ (for CI$_4$) to 24.7 kcal mol$^{-1}$ (for PbI$_4$), revealing that the σ-hole on Tt is electrophilic. The trend in the strength of the σ-hole on Tt in TtI$_4$ is in line with the polarizability of the tetrel derivative that increases in the series in this order: C(11.3 a.u.) < Si (37.3 a.u.) < Ge (40.0 a.u.) < Sn (53.0 a.u.) < Pb(56.0 a.u.) [89]. Figure 3 shows the desired relationship between them.

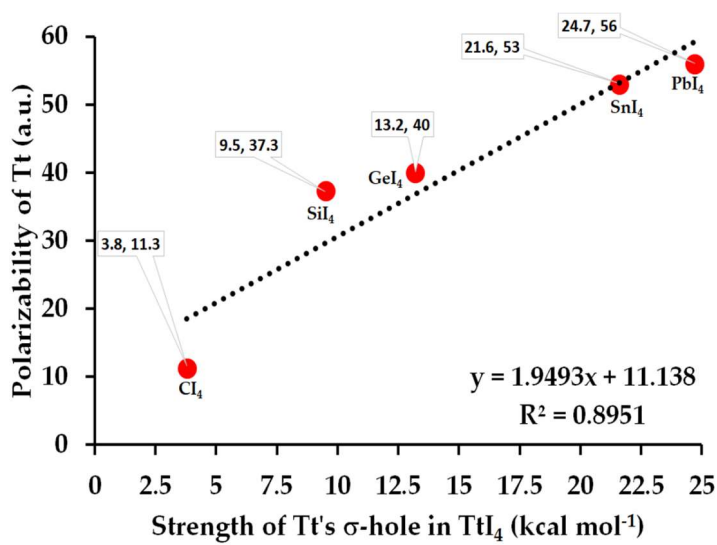

**Figure 3.** The dependence of polarizability of Tt derivative on the strength of the σ-hole on Tt in TtI$_4$, computed using [ωB97X-D/def2-QZVPPD]. The (polarizability, σ-hole) data for each molecule is indicated. The square of the regression coefficient R$^2$ is shown, together with the linear equation that connects polarizability with the strength of the σ-hole.

The strength of the σ-hole on each I atom in TtI$_4$ along the outer extension of the Tt–I covalent/coordinate bonds is also appreciable. No systematic trend in the strength of the σ-hole on each I atom is observed when passing from CI$_4$ through SiI$_4$ to GeI$_4$ to SnI$_4$ to PbI$_4$. Because the lateral portions of the covalently bonded I atoms in TtI$_4$ are equipped with negative potentials ($V_{S,min}$ values between –2.8 and –4.8 kcal mol$^{-1}$), each I atom has also have a capacity to host as a Lewis base for the attack of an electrophile. These results suggest that TtI$_4$ has the ability to function as a donor of tetrel and halogen bonds and an acceptor of tetrel bonds.

*3.2. The complex properties*

The five halide anions have linearly approached the Tt atom from the opposite side of the I–Tt covalent bond in TtI$_4$, thereby forming [I$_4$Tt···X$^-$] complexes. They are shown in Figures 4-8, in which, the Tt···X close contacts were directional (∠I–Tt···X = 180.0°).

3.2.1. The [I$_4$C···X$^-$] series

Figure 4 shows the molecular graphs of all the five molecule-anion binary complexes of CI$_4$ with the five halide anions. Because the σ-hole on the carbon atom in CI$_4$ is the weakest compared to that on the Tt atom in TtI$_4$ (Tt = Si, Ge, Sn, Pb) (Figure 2), the strength of the attractive interaction between it and the interacting halide anion(s) is weak-to-strong. For instance, the [CCSD(T)/def2-QZVPPD] level interaction energy is –3.2 and –16.35 kcal mol$^{-1}$ for [I$_4$C···F$^-$] and [I$_4$C···At$^-$] (Table 2), respectively. In all cases, the CI$_4$ unit in [I$_4$C···X$^-$], Figure 4, retains its tetrahedral shape similar to that found in its isolated counterpart (Figure 2a, Top).

The C···X intermolecular distance in [I$_4$C···X$^-$] increases as the halogen derivative becomes more polarizable; it is smallest in [I$_4$C···F$^-$], with $r$(Tt···F) = 2.665 Å with [CCSD/def2-TZVPPD]. By contrast, the I–C–I angle in complexed [I$_4$C···F$^-$] either increases or decreases compared to that of its uncomplexed counterpart (∠I–C–I = 109.47°). For instance, the I–C bond linearly attached to the anion forms smaller ∠I–C–I with the three nearest-neighbor I atoms, whereas the remaining I–C bonds that are not directly involved the formation of tetrel bond are associated with larger ∠I–C–I. These two an-

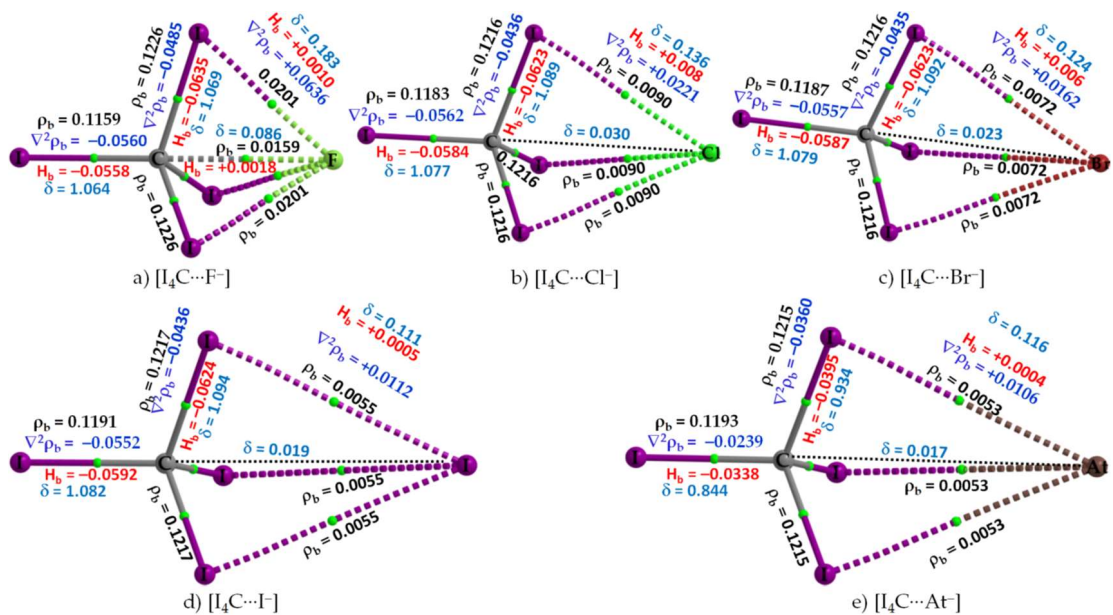

**Figure 4.** [ωB97X-D/def2-QZVPPD] level QTIM-based molecular graphs of [I$_4$C···X$^-$], showing the bond paths (solid and dotted lines in atom color) and bond critical points (tiny spheres in blue) between bonded atomic basins. Large spheres represent the atomic basins. The dotted black line in b)–e) is artificially drawn to represent the presence of tetrel bond between C and X (X = Cl, Br, I, At). The charge density($\rho_b$), the Laplacian of the charge density ($\nabla^2\rho_b$), the total energy density ($H_b$) and the delocalization index (δ) values are shown in black, blue, red and faint-blue fonts (in a.u.), respectively.

gles are 107.5 (111.3°), 108.1 (110.8°), 108.4 (110.5), 108.8 (110.1°), and 108.8 (110.1°) in [I$_4$C···F$^-$], [I$_4$C···Cl$^-$], [I$_4$C···Br$^-$], [I$_4$C···I$^-$], and [I$_4$C···At$^-$], respectively.

The bond path and bond critical point features of QTAIM appear between the C and F atomic basins indicate the presence of a C···F tetrel bond in [I$_4$C···F$^-$], Figure 4a.

The appearance of such bond path topologies between F and I atomic basins also indicates the presence of three equivalent I···F close contacts in [I4C···F−]. The latter ones mimic the I···X (X = Cl, Br, I, At) links in the other four members of the same family (Figure 4b-4e). However, QTAIM did not reveal the presence of any Tt···X tetrel bonded contacts in [I4C···X−] (X = Cl, Br, I, At). Although this kind of σ-hole interactions is expected based on inferences from the results of the MESP model (Figure 2, Bottom), their absence in the QTAIM molecular graph may be due to the stringent nature of boundaries between bonded atom basins as determined by the arbitrary nature of the space partitioning approach. Similar observations have been reported in several previous instances that do [90] or do not [91–93] involve tetrel bonding.

**Table 2.** Comparison of ωB97X-D and MP2 level intermolecular bond distances, uncorrected and BSSE corrected interaction energies of [I4Tt···F−] (Tt = C, Si, Ge, Sn, Pb; X = F, Cl, Br, I, At) with those calculated with CCSD(T). [a]

| System | [ωB97X-D/ def2-TZVPPD] | | | [ωB97X-D/ def2-QZVPPD] | | | [MP2/ def2-QZVPPD] | | | [CCSD(T)/def2-TZVPPD] | | |
|---|---|---|---|---|---|---|---|---|---|---|---|---|
|  | $E_{int}$ | $E_{int}(BSSE)$ | $r$(Tt···X) | $E_{int}$ | $E_{int}(BSSE)$ | $r$(Tt···X) | $E_{int}$ | $E_{int}(BSSE)$ | $r$(Tt···X) | $E_{int}$ | $E_{int}(BSSE)$ | $r$(Tt···X)[b] |
| [I4C···F−] | -27.36 | -19.72 | 2.690 | -19.23 | -19.2 | 2.744 | -20.11 | -18.7 | 2.663 | -26.00 | -16.35 | 2.665 |
| [I4C···Cl−] | -10.97 | -8.10 | 3.671 | -7.87 | -7.84 | 3.787 | -11.08 | -10.07 | 3.496 | -9.99 | -5.93 | 3.665 |
| [I4C···Br−] | -8.12 | -6.61 | 4.001 | -6.65 | -6.61 | 4.097 | -10.13 | -8.93 | 3.710 | -7.25 | -4.64 | 3.996 |
| [I4C···I−] | -6.02 | -5.43 | 4.422 | -5.60 | -5.57 | 4.494 | -9.25 | -7.87 | 3.990 | -4.80 | -3.38 | 4.454 |
| [I4C···At−] | -5.98 | -5.48 | 4.558 | -5.59 | -5.57 | 4.603 | -10.08 | -7.95 | 4.009 | -4.99 | -3.20 | 4.506 |
| [I4Si···F−] | -115.85 | -115.48 | 1.643 | -116.22 | -116.16 | 1.638 | -119.11 | -116.45 | 1.639 | -116.81 | -112.15 | 1.637 |
| [I4Si···Cl−] | -67.28 | -66.93 | 2.191 | -66.95 | -66.88 | 2.190 | -73.40 | -71.00 | 2.177 | -69.23 | -64.99 | 2.192 |
| [I4Si···Br−] | -9.75 | -9.61 | 3.875 | -9.85 | -9.80 | 3.875 | -62.88 | -60.00 | 2.380 | -58.1 | -53.16 | 2.408 |
| [I4Si···I−] | -7.42 | -7.36 | 4.398 | -7.52 | -7.49 | 4.398 | -53.22 | -49.92 | 2.646 | -8.51 | -6.60 | 4.287 |
| [I4Si···At−] | -7.13 | -7.08 | 4.542 | -7.23 | -7.21 | 4.539 | -13.09 | -10.87 | 3.862 | -8.44 | -6.13 | 4.397 |
| [I4Ge···F−] | -96.99 | -96.53 | 1.797 | -97.3 | -97.17 | 1.792 | -98.42 | -95.4 | 1.787 | -98.24 | -93.43 | 1.789 |
| [I4Ge···Cl−] | -60.16 | -59.75 | 2.305 | -59.77 | -59.66 | 2.305 | -65.32 | -62.64 | 2.279 | -62.93 | -58.36 | 2.300 |
| [I4Ge···Br−] | -51.15 | -50.74 | 2.505 | -50.93 | -50.8 | 2.505 | -58.29 | -55.28 | 2.456 | -54.44 | -49.32 | 2.497 |
| [I4Ge···I−] | -8.86 | -8.79 | 4.233 | -8.98 | -8.94 | 4.233 | -51.17 | -47.91 | 2.703 | -10.46 | -8.28 | 4.049 |
| [I4Ge···At−] | -8.34 | -8.28 | 4.401 | -8.44 | -8.41 | 4.401 | -51.73 | -47.2 | 2.787 | -9.95 | -7.42 | 4.203 |
| [I4Sn···F−] | -108.88 | -99.20 | 1.965 | -97.22 | -97.14 | 1.966 | -98.18 | -95.5 | 1.955 | -109.92 | -96.48 | 1.950 |
| [I4Sn···Cl−] | -72.20 | -67.68 | 2.442 | -65.22 | -65.14 | 2.447 | -70.39 | -67.92 | 2.417 | -74.18 | -65.88 | 2.427 |
| [I4Sn···Br−] | -61.88 | -59.33 | 2.751 | -58.32 | -58.21 | 2.625 | -65.06 | -62.14 | 2.576 | -63.91 | -57.29 | 2.605 |
| [I4Sn···I−] | -52.78 | -51.71 | 2.873 | -51.46 | -51.38 | 2.869 | -59.74 | -56.43 | 2.792 | -54.65 | -49.58 | 2.847 |
| [I4Sn···At−] | -51.94 | -51.05 | 2.967 | -50.56 | -50.50 | 2.966 | -60.82 | -56.21 | 2.866 | -55.43 | -49.34 | 2.928 |
| [I4Pb···F−] | -98.33 | -88.98 | 2.075 | -86.54 | -86.48 | 2.078 | -84.82 | -81.43 | 2.063 | -97.33 | -84.05 | 2.057 |
| [I4Pb···Cl−] | -67.00 | -62.66 | 2.541 | -59.88 | -59.81 | 2.548 | -63.4 | -60.33 | 2.506 | -67.71 | -59.33 | 2.521 |
| [I4Pb···Br−] | -58.51 | -56.07 | 2.715 | -54.77 | -54.67 | 2.716 | -59.91 | -56.49 | 2.655 | -59.42 | -52.72 | 2.690 |
| [I4Pb···I−] | -51.42 | -50.43 | 2.950 | -49.89 | -49.83 | 2.948 | -57.06 | -53.30 | 2.854 | -52.54 | -47.36 | 2.916 |
| [I4Pb···At−] | -51.74 | -50.91 | 3.034 | -50.20 | -50.15 | 3.032 | -59.42 | -54.41 | 2.919 | -54.51 | -48.30 | 2.984 |

[a] Interaction energies ($E_{int}$ and $E_{int}(BSSE)$) and intermolecular distances ($r$) are in kcal mol$^{-1}$ and Å, respectively. [b] Bond distances were obtained with [CCSD/def2-TZVPPD].

The authenticity of the I···X interactions in [I4C···X−] is confirmed by the I and X intermolecular distances that are close to the sum of their respective van der Waals radii (vdW), a feature which has been recommended for identifying hydrogen bonds [94], halogen bonds [95], chalcogen bond [96], pnictogen bond [35,39,62,97] tetrel bonds [27], and any other noncovalent interactions [98]. For instance, the I···F, I···Cl, I···Br, I···I and I···At intermolecular distances in [I4C···F−], [I4C···Cl−], [I4C···Br−], [I4C···I−] and [I4C···At−] are ca. 2.871, 3.626, 3.892, 4.274 and 4.322 Å, respectively. The former three are less than their respective sum of vdW radii of 3.50 (I+F), 3.86 (I+Cl), and 3.90 Å (I+Br), whereas the latter two are slightly greater than sum of their respective sum of vdW radii of 4.08 (I+I) and 4.04 Å (I+At). (vdW radii of atoms were taken from ref. [99], except for At that was taken from ref [100]). Since the vdW radii of atoms are accurate within an uncertainty of ±0.2 Å [35–37,39,46,62,99,101], the possibility of I···X (X = I, At) close contacts in [I$_4$C···I$^-$] and [I$_4$C···At$^-$] that were revealed by QTAIM are not misleading. Analogus halogen···halogen interactions in some chemical systems are known [102,103] which have been interpreted as unusually strong vdW type [103].

Both ωB97X-D and MP2 have predicted an analogous bonding scenario in [I$_4$C···X$^-$], as CCSD. However, the increase of the size of the basis set from def2-TZVPPD to def2-QZVPPD has resulted in a slight increase in the Tt···X intermolecular distance with ωB97X-D and MP2. Whatever is the size of the basis set, the intermolecular distances predicted using MP2 are underestimated relative to ωB97X-D and CCSD. Furthermore, [MP2/def2-TZVPPD] has predicted the C···I and C···At bond distances to be 4.091 and 4.083 Å for [I$_4$C···I$^-$] and [I$_4$C···At$^-$], respectively; they were 3.990 and 4.009 Å with def2-QZVPPD, respectively. This means that MP2 does not correctly predict the trend in C···I and C···At bonding distances, as predicted by DFT and CCSD (cf. Table 2).

The three I–C bonds in I$_4$C, which are not directly involved with the halide anions in [I$_4$C···X$^-$] to form the C···X tetrel bond are equivalent. Accordingly, each of the three properties such as $\rho_b$, $\nabla^2\rho_b$ ($\nabla^2\rho_b < 0$) and $H_b$ ($H_b < 0$) at the bcps of the three equivalent bonds were equivalent (one shown for each system in Figure 4). The I–C bond in I$_4$C, which is responsible for the formability of the C···X tetrel bond is largely affected in [I$_4$C···F$^-$] compared to that in the other four members of the series, and the charge density at the I–C bcp is decreased as a result of elongation of the bond. However, for all cases, all the four I–C bonds in tetrahedral I$_4$C is covalent since $H_b < 0$ and $\nabla^2\rho_b < 0$ at the I–C bcps. δ for these bonds are ranging between 0.8 and 1.1, giving evidence of their single-bond character. By contrast, $H_b > 0$ and $\nabla^2\rho_b > 0$ for the C···F bcp in [I$_4$C···F$^-$], and I···X bcps in [I$_4$C···X$^-$], and the charge density is very small at corresponding bcps. Moreover, δ for the atom-atom pairs responsible for C···X decrease systematically in the series in this order: [I$_4$C···F$^-$] (δ = 0.086) > [I$_4$C···Cl$^-$] (δ = 0.030) > [I$_4$C···Br$^-$] (δ = 0.023) > [I$_4$C···I$^-$] (δ = 0.019) > [I$_4$C···At$^-$] (δ = 0.017), in consistent with the trend found for interaction energy (Table 2).

3.2.2. The [I$_4$Tt···X$^-$] (Tt = Si, Ge) series

The nature of the intermolecular bonding environment found in [I$_4$C···X$^-$] is not the same for all the five members of the series [I$_4$Si···X$^-$] or [I$_4$Ge···X$^-$] (cf. Figure 5 and 6, respectively). Because the electrostatic surfaces of Si and Ge in SiI$_4$ and GeI$_4$, respectively, were relatively more electrophilic than that on C in CI$_4$, they showed reasonably strong selectivity for the anions. This was specifically true when X pointed to F, and Cl (Figure 5a-b), but not when X pointed to Br, I, and At in [I$_4$Si···X$^-$] (Figure 5c-e). Similarly, the Ge atom in GeI$_4$ appreciably recognizes the halide anions when X is F, Cl, or Br (Figure 6a-c), but not when X is I, or At in [I$_4$Ge···X$^-$] (Figure 6d-e). This means that the strength of Tt···X bonding is moderate when the latter two heavy halide anions are involved, in which, the degree of tetrahedral deformation of the I$_4$Si/I$_4$Ge unit in [I$_4$Si···X$^-$] /[I$_4$Ge···X$^-$] is small. When the I$_4$Si/I$_4$Ge isolated monomers strongly recognizes the halide anions, the degree of deformation of I$_4$Si/I$_4$Ge is overwhelmingly large, and hence the tetrahedral shape of SiI$_4$/GeI$_4$ in [I$_4$Si···X$^-$]/ [I$_4$Ge···X$^-$] (X = F, Cl, or Br) is completely lost.

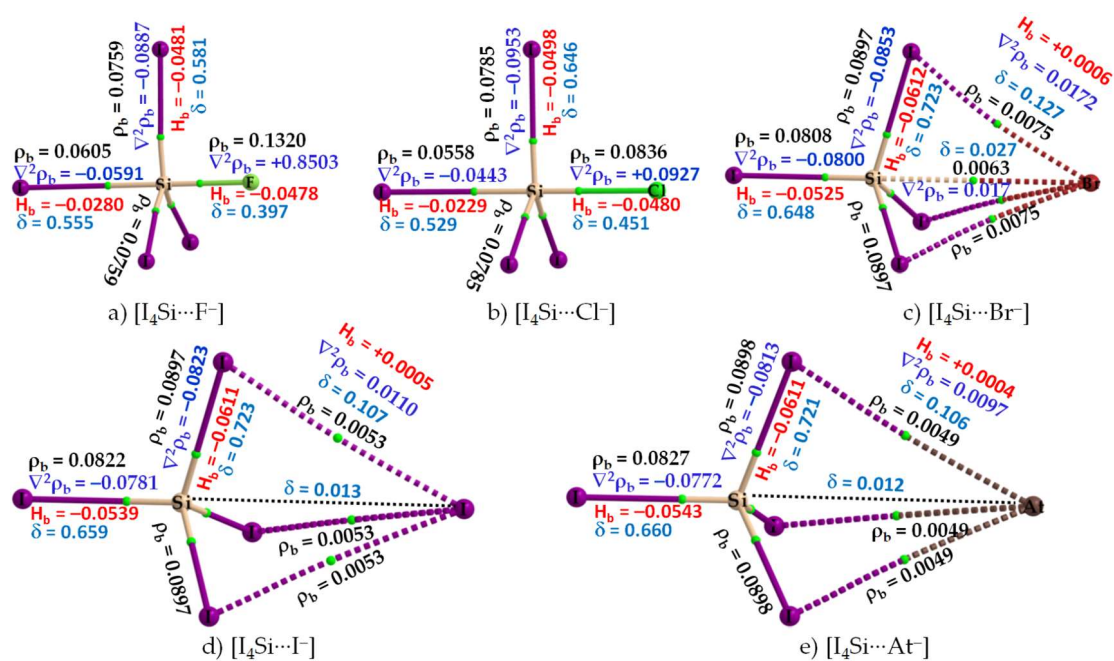

**Figure 5.** [ωB97X-D/def2-QZVPPD] level QTIM-based molecular graphs of [I$_4$Si⋯X$^-$] (X = F, Cl, Br, I, At), showing the bond paths (solid and dotted lines in atom color) and bond critical points (tiny spheres in blue) between bonded atomic basins. Large spheres represent the atomic basins, with atoms labelled. The dotted black line in d)–e) is artificially drawn to represent the presence of tetrel bond between Si and X (X = I, At). The charge density($\rho_b$), the Laplacian of the charge density ($\nabla^2\rho_b$), the total energy density (H$_b$) and the delocalization index (δ) values are shown in black, blue, red and faint-blue fonts (in a.u.), respectively.

While the bonding features noted above were obtained from [ωB97X-D/def2-QZVPPD], the [CCSD/def2-QZVPPD] method has predicted that the interacting units responsible for [I$_4$Si⋯Br$^-$] should involve in the formation of a dative tetrel bond; it should be in a manner similar to that found for [I$_4$Si⋯X$^-$] (X = F, Cl). On the other hand, MP2 has recognized the attraction between I$_4$Si and X$^-$ in the first four members of the [I$_4$Si⋯X$^-$] series to be unusually strong, and that in [I$_4$Si⋯At$^-$] to be moderate. The former observation with MP2 is applicable to the [I$_4$Ge⋯X$^-$] series as well. This means that the Tt⋯X close contacts in these molecule-anion systems are not ordinary tetrel bonds, they are dative tetrel bonds.

QTAIM analysis, Figure 6a-c, revealed that $\rho_b$ is appreciable at the Ge⋯X bcps in [I$_4$Ge⋯X$^-$] when X points to F, Cl and Br. For [I$_4$Ge⋯I$^-$], the $\rho_b$ is small at Ge⋯I bcp ($\rho_b$ = 0.0048 a.u.), and the interaction between the monomers is reinforced by I⋯I interactions (Figure 6d). The $\rho_b$ values at the Ge⋯X (X = F, Cl, Br) bcps in [I$_4$Ge⋯X$^-$] are not only typical for coordinate bonds but larger than that can be expected for ordinary non-covalent interactions such as hydrogen bonds, and halogen bonds, among others ($\rho_b$ < 0.05 a.u.). They may be comparable with the $\rho_b$ values of the Tt—I coordinate bonds in isolated and complexed TtI$_4$. A similar conclusion might be arrived at for Si⋯X bcps in [I$_4$Si⋯X$^-$] (X = F, Cl), Figure 5a-b.

From the sign and magnitude of $\nabla^2\rho_b$ ($\nabla^2\rho_b$ > 0) and H$_b$ (H$_b$ < 0), Figure 6a-c, it is realized that the Ge⋯X (X = F, Cl, Br) tetrel bonds possess mixed ionic and covalent character. This view is also transferable to the Si⋯X (X = F, Cl, Br) tetrel bonds in [I$_4$Si⋯X$^-$] provided [CCSD/def2-TZVPPD] results are considered. The large δ values corresponding to atom-atom pairs responsible for the Si⋯X and Ge⋯X (X = F, Cl, Br) contacts provide further evidence that there are no π-type interactions involved; they are purely σ-type coordinate dative bonds. By contrast, the δ values are very small for atom-atom pairs causing the Si⋯X and Si⋯X close contacts in [I$_4$Ge⋯X$^-$] and [I$_4$Ge⋯X$^-$] (X = I, At), respectively, indicative of closed-shell interactions. The three equivalent I···X close contacts in [I$_4$Ge···X$^-$] and [I$_4$Ge···X$^-$] (X = I, At) are described by small δ values, and positive ∇$^2$ρ$_b$ and H$_b$. Similarly, the Si···X and Ge···X (X = I, At) tetrel bonds are described by small δ values, as expected.

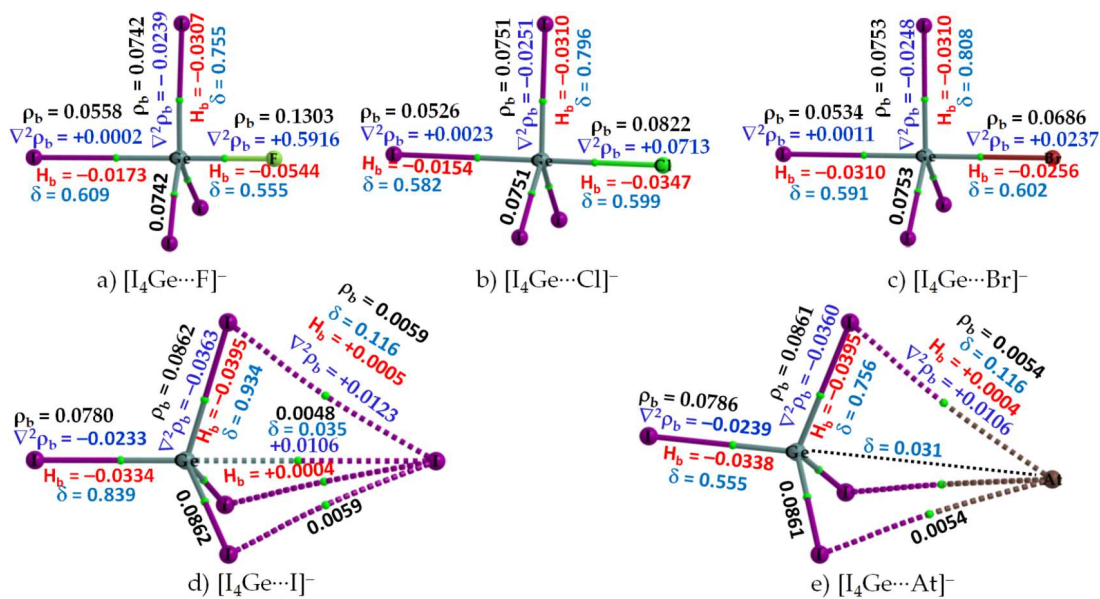

**Figure 6.** [ωB97X-D/def2-QZVPPD] level QTIM-based molecular graphs of [I$_4$Ge···X$^-$] (X = F, Cl, Br, I, At), showing the bond paths (solid and dotted lines in atom color) and bond critical points (tiny spheres in blue) between bonded atomic basins. Large spheres represent the atomic basins, with atoms labelled. The dotted black line in e) is artificially drawn to represent the presence of tetrel bond between Ge and At. The charge density (ρ$_b$), the Laplacian of the charge density (∇$^2$ρ$_b$), the total energy density (H$_b$) and the delocalization index (δ) values are shown in black, blue, red and faint-blue fonts (in a.u.), respectively.

3.2.3. The [I$_4$Tt···X$^-$] (Tt = Sn, Pb) series

The σ-holes on Sn and Pb in SnI$_4$ and PbI$_4$, respectively, are stronger than those of TtI$_4$ (Tt = C, Si, Ge). Therefore, their acidic strengths are adequately enough to recognize the five halide anions when in close proximity. This may be rationalized from QTAIM's molecular graphs of resulting configurations, [I$_4$Sn···X$^-$] and [I$_4$Pb···X$^-$] (X = F, Cl, Br, I, At), illustrated in Figures 7 and 8, respectively. As can be seen, the formation of an intermolecular interaction in each of them has caused profound damage to tetrahedral framework of isolated SnI$_4$ and PbI$_4$. This means that the TtI$_4$ molecule is structurally fully deformed in the presence of each of the five halide anions. There is no secondary intermolecular interaction that can play a role in the stability of the resulting complex anion, as found for other series (see above). In all cases, the tetrel center adopts to a trigonal bipyramidal geometry (a molecular structure with one atom at the center and five more atoms at the corners of the trigonal bipyramid). Clearly, the resulting complex anions each is nothing but a coordination compound, and the Tt···X close-contact is formally a Tt—X dative tetrel bond. In such cases, the degree of charge transfer from the anion to the σ*(I–Tt) anti-bonding orbital is expected. The S$_N$2 mechanism is likely to play a role in driving the reaction [23].

Our calculation suggests that the extent of charge transfer is the largest for the [I$_4$Pb···X$^-$] series and the smallest for the [I$_4$C···X$^-$] series. In particular, the [ωB97X-D/def2-QZVPPD] level QTAIM charge transfer from X$^-$ to PbI$_4$ is 0.263, 0.371, 0.424, 0.515 and 0.588 *e* for [I$_4$Pb···F$^-$], [I$_4$Pb···Cl$^-$], [I$_4$Pb···Br$^-$], [I$_4$Pb···I$^-$], and [I$_4$Pb···At$^-$], respectively.

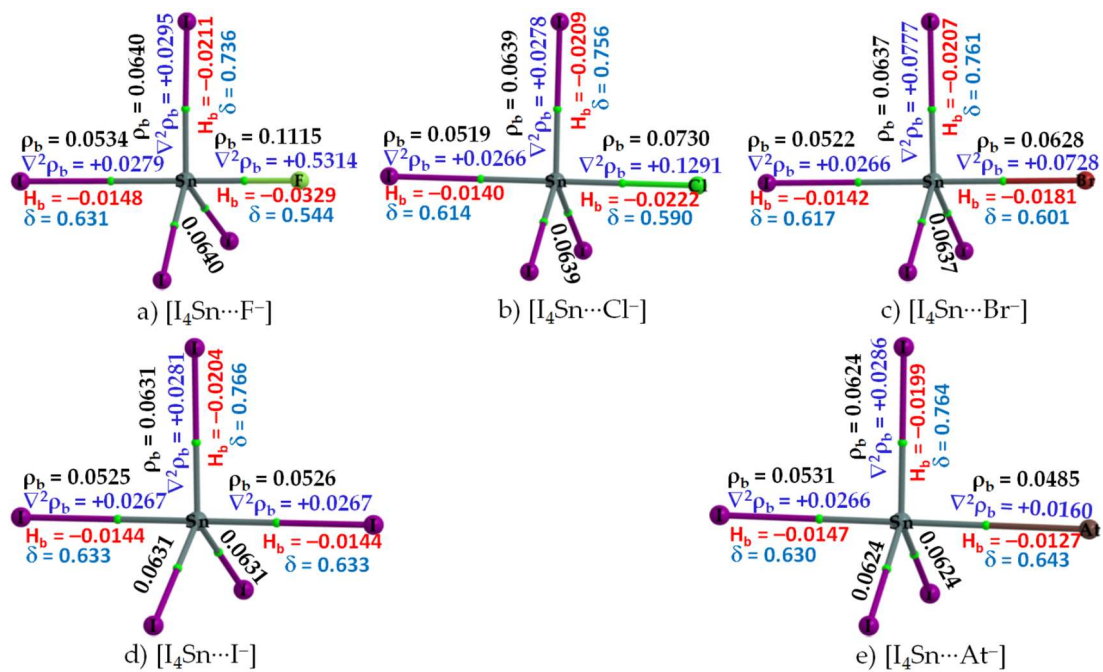

**Figure 7.** [ωB97X-D/def2-QZVPPD] level QTIM-based molecular graphs of [I$_4$Sn···X$^-$] (X = F, Cl, Br, I, At), showing the bond paths (solid and dotted lines in atom color) and bond critical points (tiny spheres in blue) between bonded atomic basins. Large spheres represent the atomic basins, with atoms labelled. The charge density ($\rho_b$), the Laplacian of the charge density ($\nabla^2\rho_b$), the total energy density ($H_b$) and the delocalization index ($\delta$) values are shown in black, blue, red and faint-blue fonts (in a.u.), respectively.

The corresponding charge transfer values were 0.229, 0.334, 0.382, 0.460 and 0.519 $e$ for [I$_4$Sn···F$^-$], [I$_4$Sn···Cl$^-$], [I$_4$Sn···Br$^-$], [I$_4$Sn···I$^-$], and [I$_4$Sn···At$^-$], respectively. These results imply that the nature of charge-transfer in the Sn- and Pb-based anions is virtually similar and that the charge-transfer preference is consistent with the interaction energy preference across a given series, indicating that the charge-transfer phenomenon is likely to be one of the most prominent contributors to the interaction. The nature of charge transfer noted above is moderately large, for example, relative to that of 0.115, 0.088, 0.082, 0.077 and 0.076 $e$ calculated for [I$_4$C···F$^-$], [I$_4$C···Cl$^-$], [I$_4$C···Br$^-$], [I$_4$C···I$^-$], and [I$_4$C···At$^-$], respectively. These results lead to an observation that stronger complexes are accompanied with stronger transfer of charge between the interacting monomers, which is not very surprising [104].

The molecular graphs of QTAIM in Figures 7 and 8 show that the charge density at Tt···X bcps between I$_4$Tt and X$^-$ in [I$_4$Tt···X$^-$] (Tt = Sn, Pb; X = F, Cl, Br, I, At) is non-negligible; it may be comparable to the Tt—I bcps of complexed I$_4$Tt; the Tt—I and Tt···X bcps are both characterized by $\nabla^2\rho_b > 0$ and $H_b < 0$, indicating the presence of a mixed covalent and ionic character. Since $H_b$ becomes increasingly more positive at the Tt···X bcp passing from [I$_4$Tt···F$^-$] through [I$_4$Tt···Cl$^-$] to [I$_4$Tt···Cl$^-$] to [I$_4$Tt···Br$^-$] to [I$_4$Tt···I$^-$] to [I$_4$Tt···At$^-$], it is clear that these interactions are less covalent in the same order. The $\nabla^2\rho_b$ values become progressively smaller in the series from [I$_4$Tt···F$^-$] through [I$_4$Tt···Cl$^-$] to [I$_4$Tt···At$^-$], indicating that Tt···F is more ionic than Tt···At. In the case of Tt = Si and Ge, the Tt···X (X = F, Cl, Br) bcps show $\nabla^2\rho_b > 0$ and $H_b < 0$ (Figures 5 and 6). The four Si—I bonds in I$_4$Si of [I$_4$Si···X$^-$] are potentially covalent since $\nabla^2\rho_b < 0$ and $H_b < 0$ at the bcps of these bonds, as like as the three Ge—I bonds in I$_4$Ge of [I$_4$Ge···X$^-$] that are orthogonal to the Tt···X (X = F, Cl, Br) bond in the respective system. The characteristics of Si—I bonds in I$_4$Si resemble the C—I bonds in I$_4$C.

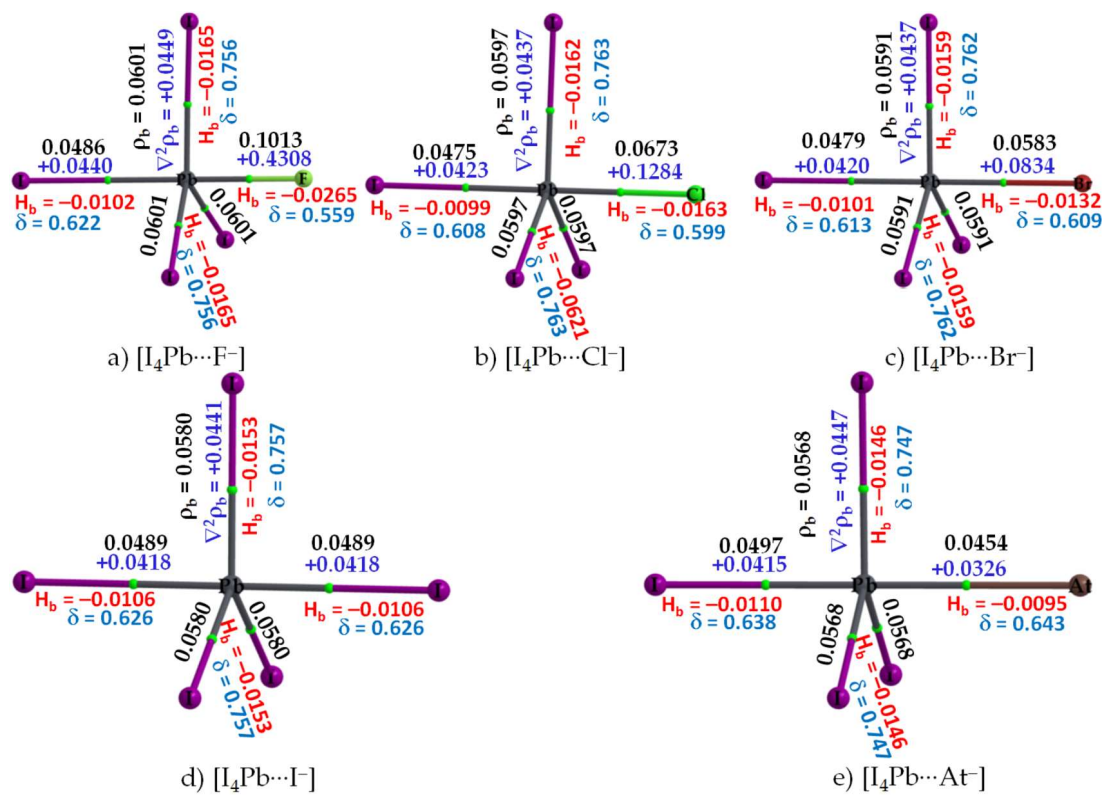

**Figure 8.** [ωB97X-D/def2-QZVPPD] level QTIM-based molecular graphs of [I₄Pb···X⁻] (X = F, Cl, Br, I, At), showing the bond paths (solid and dotted lines in atom color) and bond critical points (tiny spheres in blue) between bonded atomic basins. Large spheres represent the atomic basins, with atoms labelled. The charge density (ρ_b), the Laplacian of the charge density (∇²ρ_b), the total energy density (H_b) and the delocalization index (δ) values are shown in black, blue, red and faint-blue fonts (in a.u.), respectively.

The δ of the atom-atom pairs for Tt—I and Tt···X bonds in [I₄Tt···X⁻] (Tt = Sn, Pb) is considerably larger than what were calculated for ordinary tetrel bonds (see Figure 6a-c for the former and Figure 6d-e for the latter bonds, for example). It is considerably smaller than those in isolated I₄Tt (Figure 2, Top), thus consistent with the weakening of the Tt—I bond in I₄Tt (Tt = Si, Ge, Sn, Pb) upon its attractive engagement with the halide anions. For comparison, we note that the δ values for Tt—I and Tt···X coordinate and dative tetrel bonds in [I₄Tt···X⁻] are smaller than, and comparable to, those reported for metal—C(O) coordinate bonds (δ values ranged from 0.279 to 1.195); however, those of Tt···X ordinary tetrel bonds are comparable with what were reported for metal···metal (metal···H or H···H) interactions (δ values between 0.005 and 0.166) in [M₂(CO)₁₀] and [M₃(μ-H)₃(CO)₁₂] (M = Mn, Tc, Re) complexes [105].

3.2.4. IGM-δg^inter analysis

The formation of [I₄Tt···X]⁻ has caused weakening of all the four Tt—I bonds in TtI₄, compared to that found in the uncomplexed TtI₄. The weakening was evidence of the elongation of the Tt—I bonds in [I₄Tt···X]⁻. Concomitant with the elongation was the decrease in the charge density at the Tt—I bcps, which may be inferred comparing the ρ_b values at the Tt—I bcps shown in Figure 2 (Top) for isolated TtI₄ and in Figure 4-8 for complexed TtI₄. Th existence of I···X and Tt···X in some complexes of [I₄Tt···X]⁻ (Tt = C, Si and Ge) are also genuine, which are confirmed by the results of IGM-δg^inter analysis.

Figure 9 illustrates the results of IGM-δg^inter analysis for [I₄Tt···X⁻] (Tt = C). The I···X closed contacts in several of these systems appeared at larger IGM-δg^inter isovalues (Top). On the other hand, the Tt···X close contacts showed up at lower IGM-δg^inter isovalues

(Bottom). These features are not very surprising since smaller isolvalues are typically necessary for the physical appearances of isosurfaces corresponding to weakly bonded interactions. By contract, the relatively stronger interactions can be traceable with larger isovalues since charge density around critical bonding region is appreciable. The bluish isosurface, originated with large IGM-$\delta g^{inter}$ isovalues, for [I$_4$C···F$^-$] indicates that the attraction between the interacting units is very prominent. When the size of the halogen derivative increases, the attraction between C and X in [I$_4$C···X$^-$] weakens and hence the isosurfaces becomes increasingly greenish. These results are concordant with the nature of the QTAIM-based charge density features at the I···F and C···F bcps (cf. Figure 4). Therefore, the stabilization of [I$_4$C···X$^-$] is not just due to the formation of the C···X tetrel bonds alone, but partly arises from the I···X Type-I halogen···halogen bonded interactions as well.

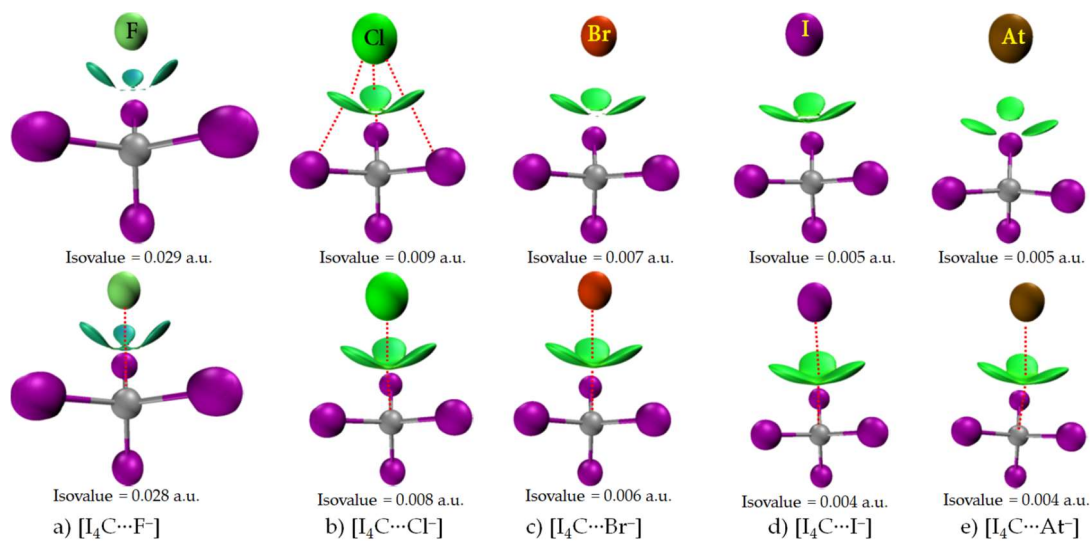

**Figure 9.** (a)-(e) [ωB97X-D/def2-QZVPPD] level IGM-$\delta g^{inter}$ based isosurface (bluish-green or green volumes) plots of [I$_4$C···X$^-$] (X = F, Cl, Br, I, At), showing possible I···X halogen bonded and C···X tetrel bonded interactions between interacting molecular entities. (Top) Illustration of I···X Type-I halogen-halogen bonded interactions between the interacting units that appear with larger isovalues. (Bottom) Illustration of C···X tetrel bonded interactions between the interacting units that appear at smaller isovalues. Anion derivatives are labelled.

The [I$_4$Si···X$^-$] systems feature very similar IGM-$\delta g^{inter}$ based isosurfaces, Figure 10. That is, a very large isovalue was necessary to reveal sizable isosufaces describing the dative tetrel bond in [I$_4$Si···F$^-$] and [I$_4$Si···Cl$^-$], whereas a potentially small isovalue was required to visualize the isosurface domains in [I$_4$Si···X$^-$] (X = Br, I, At). The view is also transferable to the [I$_4$Ge···X$^-$] systems (not shown).

In the case of [I$_4$Sn···X$^-$] and [I$_4$Pb···X$^-$] (X = F, Cl, Br, I, At), the IGM-$\delta g^{inter}$ isosurfaces were visualizable with an isovalue close to 0.055 a.u. Figure 11 shows this for the [I$_4$Pb···X$^-$] series. Passing from the left to the right of Figure 11, it can be seen that the thickness and size of the bluish isosurface volume describing the tetrel bond between Tt and X are decreasing. This is also in agreement with QTAIM in that the charge density between these atomic basins decreases from [I$_4$Tt···F$^-$] through [I$_4$Tt···Cl$^-$] to [I$_4$Tt···Br$^-$] to [I$_4$Tt···I$^-$] to [I$_4$Tt···At$^-$] (Tt = Sn, Pb).

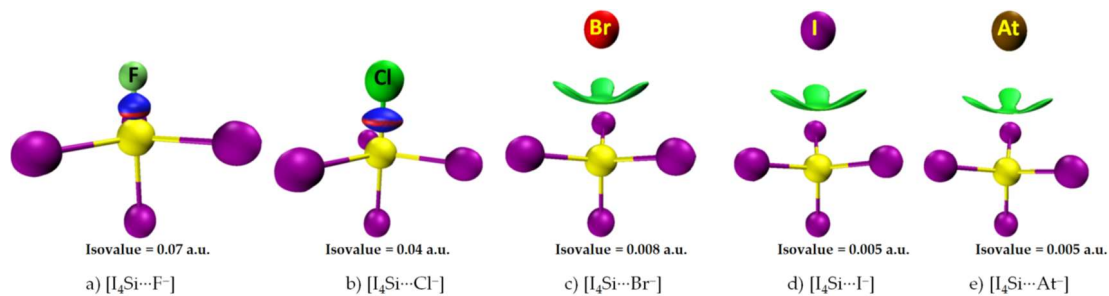

**Figure 10.** (a)-(e) [ωB97X-D/def2-QZVPPD] level IGM-$\delta g^{inter}$ based isosurface (bluish-red or green volumes) plots for [I$_4$Si⋯X$^-$] (X = F, Cl, Br, I, At), showing possible I⋯X halogen-halogen bonded and Si⋯X tetrel bonded interactions between interacting molecular entities.

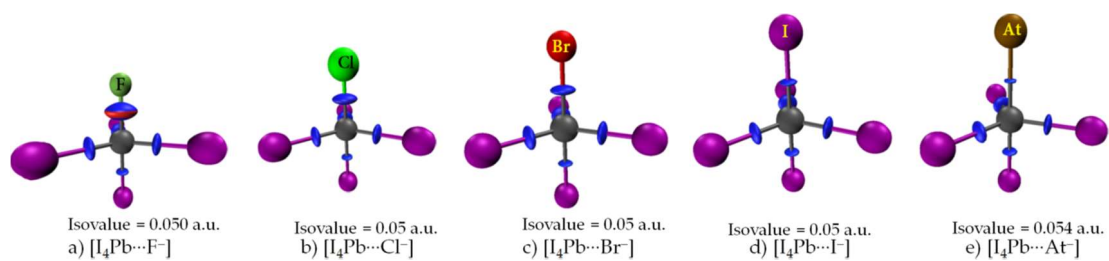

**Figure 11.** (a)-(e) [ωB97X-D/def2-QZVPPD] level IGM-$\delta g^{inter}$ based isosurface (bluish-red or green volumes) plots of [I$_4$Pb⋯X$^-$] (X = F, Cl, Br, I, At), showing possible Pb⋯X tetrel bonded interactions between interacting molecular entities.

### 3.2.2. Interaction energies

Except for the [I$_4$C⋯X$^-$] series, and [I$_4$Tt⋯X$^-$] (Tt = Si, Sn; X = I, At), the $E_{int}$ (BSSE) values for all other molecule-anion systems are much larger than the so-called covalent limit for hydrogen bonds (–40.0 kcal mol$^{-1}$) [106–108]. From the values of the interaction energies of 25 molecule-anion complexes, [I$_4$Tt⋯X$^-$], it is clear that the complex stability is largely determined by the polarizability of the Tt atom in I$_4$Tt and the halogen derivative. These energies calculated in the range from –3.0 and –112.2 kcal mol$^{-1}$ with [CCSD(T)/def2-TZVPPD], Table 2, can be categorized as weak (–3.0 kcal mol$^{-1}$ < $E_{int}$ (BSSE) < –5.0 kcal mol$^{-1}$), moderate (–5.0 kcal mol$^{-1}$ < $E_{int}$ (BSSE) < –10.0 kcal mol$^{-1}$), strong (–10.0 kcal mol$^{-1}$ < $E_{int}$ (BSSE) < –25.0 kcal mol$^{-1}$), very strong (–25.0 kcal mol$^{-1}$ < $E_{int}$ (BSSE) ≤ –40.0 kcal mol$^{-1}$), and ultra-strong ($E_{int}$ (BSSE) >> –40.0 kcal mol$^{-1}$ (covalent limit for hydrogen bond)). At the highest level of theory applied, [CCSD(T)/def2-TZVPPD], the weakest and strongest of the [I$_4$Tt⋯X$^-$] systems are found to be [I$_4$C⋯At$^-$] ($E_{int}$ (BSSE) = –3.20 kcal mol$^{-1}$) and [I$_4$Si⋯F$^-$] ($E_{int}$ (BSSE) = –112.15 kcal mol$^{-1}$), respectively.

From Table 2, two major differences are noteworthy. First, ωB97X-D predicts a BSSE-corrected interaction energy of –9.61 and –9.80 kcal mol$^{-1}$ for [I$_4$Si⋯Br$^-$] with def2-TZVPPD and def2-QZVPPD, respectively; these are indicative of the fact that the strength of the tetrel bond between Si of I$_4$Si and Br$^-$ is moderate. As mentioned already above, this is not the case with MP2 since the $E_{int}$ (BSSE) for the same system, for instance, with def2-QZVPPD, is predicted to be –60.0 kcal mol$^{-1}$; the large $E_{int}$ (BSSE) implies that the attraction between I$_4$Si and Br$^-$ causes the formation of Si—Br dative tetrel bond. This result is consistent with [CCSD(T)/def2-TZVPPD], which has predicted an $E_{int}$ (BSSE) of –53.16 kcal mol$^{-1}$ for the same system. Second, the [ωB97X-D/ def2-TZVPPD] level $E_{int}$ (BSSE) values for the remaining four systems of the [I$_4$Si⋯X]$^-$ series are in qualitative and quantitative agreement with [CCSD(T)/def2-TZVPPD]. MP2, however, unusually overestimated the interaction energies for [I$_4$Si⋯I$^-$] and [I$_4$Si⋯At$^-$]. The discrepan-cy between the DFT (or CCSD(T)) and MP2 energies is likely due to the latter method's misleading prediction of the Si···I and Si···At close-contacts, thus pushing the interacting atoms in [I$_4$Si···I$^-$] to be bonded with each other via a dative tetrel bond. These peculiar results indicate that applying the MP2 approach to predict the correct nature of the tetrel bond in molecule-anion complex systems formed by heavier tetrel derivatives in molecular entities should be exercised with caution.

The preference of BSSE-corrected interaction energy, $E_{int}$ (BSSE), between the five members of each series [I$_4$Tt···X]$^-$ follows the trend: [I$_4$Tt···F$^-$] > [I$_4$Tt···Cl$^-$] > [I$_4$Tt···Br$^-$] > [I$_4$Tt···I$^-$] > [I$_4$Tt···At$^-$] (Table 2). This is the energy preference at the highest level of theory applied, [CCSD(T)/def2-TZVPPD], which shows a tendency for the strength of the interaction to decrease with increasing polarizability of the halogen derivative (F < Cl < Br < I < At). This stability preference could not be reproduced with ωB97X-D when def2-TZVPPD was used since it altered the stability priority between [I$_4$Tt···I$^-$] and [I$_4$Tt···At$^-$] when Tt = C and Pb, giving rise to: [I$_4$Tt···F$^-$] > [I$_4$Tt···Cl$^-$] > [I$_4$Tt···Br$^-$] > [I$_4$Tt···I$^-$] ≤ [I$_4$Tt···At$^-$]. The same trend was also observed when MP2 was used in conjunction with def2-NZVPPD (N = T, Q). Note that changing the basis set from def2-TZVPPD to def2-QZVPPD somehow restored the [CCSD(T)/def2-TZVPPD] level energy preference at the ωB97X-D level (but not with MP2) when Tt = C, but not when Tt = Pb. One reason for the anomalous change in the preference of energy ordering between [I$_4$Tt···I$^-$] and [I$_4$Tt···At$^-$] is that the post-HF MP2 method greatly overestimates the BSSE, as well the electron-electron correlation energy, relative to the DFT and CCSD(T). On the other hand, the CCSD(T) method has properly accounted for electron-electron correlation energy, which ensured the correct preference of stabilization energies among the five members of any given [I$_4$Tt···X$^-$] series.

Figure 12a-c compares the type of dependence of $E_{int}$(BSSE) on the distance of separation $r$(Tt···X) for 25 molecule-anion complexes, [I$_4$Tt···X$^-$], obtained using ωB97X-D, MP2 and CCSD(T). Regardless of the different calculation methods utilized, the dependence was found to be quadratic. The square of the regression coefficient $R^2$ was moderately higher ($R^2$=0.9325) for ωB97X-D compared to CCSD(T) and lower ($R^2$=0.8923) for MP2.

We note further that the BSSE in energy is minimal with DFT but larger with MP2 and CCSD(T). It is very large with the def2-TZVPPD basis set than with the def2-QZVPPD basis set. For example, for ωB97X-D, MP2, and CCSD(T) with def2-TZVPPD, the BSSE in energy ranged from 0.05 to 9.68 kcal mol$^{-1}$, from 2.60 to 13.58 kcal mol$^{-1}$, and from 1.42 to 13.44, respectively. However, when using the def2-QZVPPD basis set, the BSSE in energy has decreased sharply, giving rise to values in the range from 0.02 to 0.13 kcal mol$^{-1}$ with ωB97X-D and from 1.01 to 5.01 kcal mol$^{-1}$ with MP2. CCSD(T) with def2-QZVPPD was computationally very expensive, no conclusions could be drawn about the range of BSSE in energy with this method. Figure 12d-f compares the nature of dependence between $E_{int}$(BSSE) and $E_{int}$, obtained using [ωB97X-D/def2-QZVPPD], [MP2/def2-QZVPPD] and [CCSD(T)/def2-TZVPPD], respectively, showing a perfect linear dependence at the former level than that at the latter two.

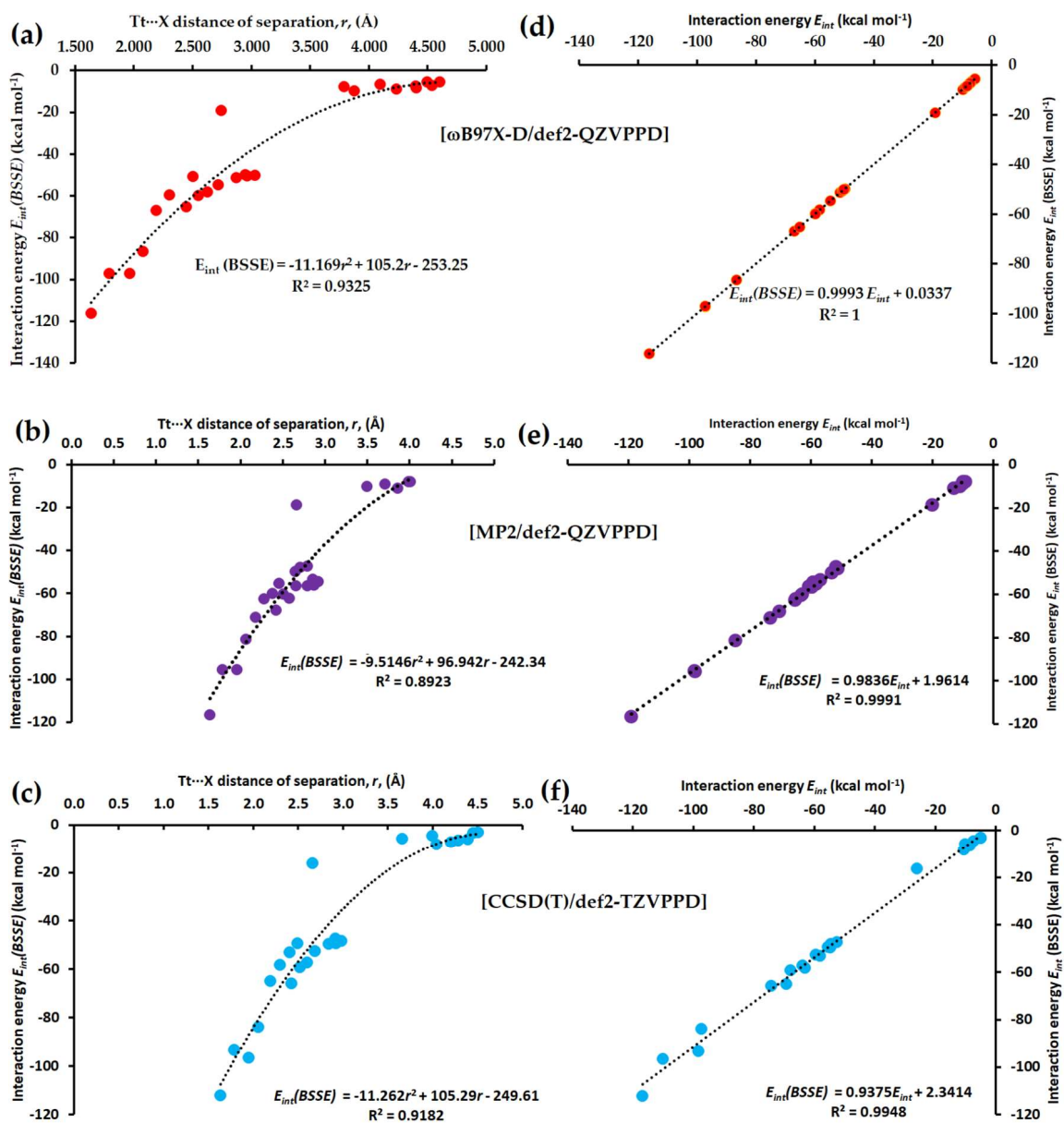

**Figure 12:** (a) The [ωB97X-D/def2-QZVPPD] level quadratic dependence of BSSE corrected interaction energy ($E_{int}(BSSE)$) on the distance of separation $r$(Tt···X) for 25 [I$_4$Tt···X$^-$] (Tt = C, Si, Ge, Sn, Pb; X = F, Cl, Br, I, At) molecule-anion complexes. Shown in (b) and (c) are the corresponding dependences obtained using [MP2/def2-QZVPPD] and [CCSD(T)/def2-TZVPPD], respectively. Shown in (e), (f) and (g) are the linear dependences between $E_{int}(BSSE)$ and $E_{int}$ obtained with [ωB97X-D/def2-QZVPPD], [MP2/def2-QZVPPD] and [CCSD(T)/def2-TZVPPD], respectively.

## 4. Discussion and Concluding Remarks

In this study, the series [I$_4$Tt···X$^-$] (Tt = C, Si, Ge, Sn, Pb; X = F, Cl, Br, I, At) was theoretically investigated to clarify the nature of the selectivity of the I$_4$Tt host for five guest anions. The MP2 geometries and interaction energies for the 25 molecule-anion systems were underestimated and overestimated, respectively, relative to DFT and CCSD methods, and in some cases the MP2 results were unreliable. The chemical bonding features obtained using DFT were consistent with the computationally expensive CCSD and CCSD(T) results, with an exception for [I$_4$Si···Br$^-$]. For the latter, the tetrel bonding characteristics predicted by CCSD could not be reproduced by DFT-ωB97X-D. Similarly, the significant overestimation of the interaction energy of [I$_4$Si···I$^-$] with MP2 was in sharp disagreement with ωB97X-D and CCSD(T).

The deformation of the tetrahedral skeletal framework of TtX$_4$ was shown to be prominent especially when the electron-withdrawing anions, viz. F$^-$ and Cl$^-$, and some-times Br⁻, were used as partner interacting species for TtI$_4$ (Tt = Si, Ge, and Sn). When Sn and Pb of TtI$_4$ were acted as tetrel bond donors for all the five anions, the original tetrahedral shape of parent TtI$_4$ was completely lost, and the tetrel atoms in the resulting complex anion systems preferred adopting a pentagonal bipyramidal geometry. This was attributed to the strong electrophilicity (greater selectivity) of the heavier tetrel derivatives, in which, appreciable charge transfer occurred from the anion to the tetrel donor moiety that led to dative tetrel bond formation.

In several complexes, tetrel bonding did not occur alone. This was true especially when the molecule hosting the tetrel bond donor was not fully deformed. That is, the same anion that caused the formation of an ordinary Tt···X tetrel bond was simultaneously engaged with three nearest-neighbor iodine atoms that were responsible for an I$_3$ face of the TtI$_4$ tetrahedron. In such cases, the strength of the Tt···X tetrel bond might be reinforced by the I···X interactions, evidenced by the charge-density based topological results of QTAIM and IGM-$\delta g^{inter}$. Therefore, these results enable us to believe that a similar intermolecular bonding scenario might be existing between the interacting monomers responsible for some of the systems of the series [Y$_4$Tt···X⁻] (Tt = C, Si, Ge; Y = F, Cl Br) [21]; further computational studies on them are a perquisite to validate our claim.

The Tt···X separation distance calculated by [CCSD/def2-TZVPPD] was smaller than the vdW radii of the Tt and X atoms for all [I$_4$Tt···X⁻] systems, except for [I$_4$C···X⁻] (X = Cl, Br, I, At) and [I$_4$Si···X⁻] (X = I, At). This was the case with [ωB97X-D/def2-TZVPPD] and [ωB97X-D/def2-QZVPPD], but [I$_4$Ge···At⁻] was added to the exclusion list. The result was different from [MP2/def2-QZVPPD] in that it predicted exceptions only for [I$_4$C···X⁻] (X = Br, I, At), and is not surprising given it is an MP2's tendency to underestimate intermolecular distances. Although these latter two computational methods exclude intermolecular interactions in systems that do not follow the stringent "less than the sum of vdW radii rule," the exclusion was also consistent with the bond path topology of QTAIM. For example, for [I$_4$C···X⁻] (X = Cl, Br, I, At), [I$_4$Tt···X⁻] (Tt = Si; X = I, At) and [I$_4$Tt···X⁻] (Tt = Ge; X = At), no QTAIM-based bond path topology exists between Tt and X at the [ωB97X-D/def2-QZVPPD] level. This means that QTAIM does not recognize the existence of Tt···X tetrel bonding in the host-guest systems when the tetrel bond distance between Tt and X exceeded the sum of the vdW radii of Tt and X, even though this type of limitation of QTAIM has been attributed to the arbitrary nature of the space partitioning scheme. However, the Tt and X atoms of the interacting monomers in all systems were indeed tetrel bonded to each other, evidenced by the IGM-$\delta g^{inter}$ based isosurfaces between bonded atomic basins.

**Funding:** This research received no external funding.

**Institutional Review Board Statement:** Not applicable.

**Informed Consent Statement:** Not applicable.

**Data Availability Statement:** This research did not report any data.

**Acknowledgments:** This work was entirely conducted using the various laboratory facilities provided by the University of Tokyo and University of the Witwatersrand. The author is currently affiliated with University of the Witwatersrand (SA) and Nagoya University, Aichi 464-0814, Japan.

**Conflicts of Interest:** The author declares no conflict of interest. The funders had absolutely no role in the design of the study; in the collection, analyses, or interpretation of data; in the writing of the manuscript; or in the decision to publish the results.

**References**

1. Sethio, D.; Raggi, G.; Lindh, R.; Erdélyi, M. Halogen Bond of Halonium Ions: Benchmarking DFT Methods for the Description of NMR Chemical Shifts. *J. Chem. Theory Comput.* **2020**, *16*, 7690–7701, doi:10.1021/acs.jctc.0c00860.

2. Chen, Y.; Wang, D.-X.; Huang, Z.-T.; Wang, M.-X. Ion Pair Receptors Based on Anion–π Interaction. *Chem. Commun.* **2011**, *47*, 8112–8114, doi:10.1039/C1CC12075A.

3. Beer, P.D.; Gale, P.A. Anion Recognition and Sensing: The State of the Art and Future Perspectives. *Angewandte Chemie International Edition* **2001**, *40*, 486–516, doi:10.1002/1521-3773(20010202)40:3<486::AID-ANIE486>3.0.CO;2-P.

4. Lim, J.Y.C.; Beer, P.D. Sigma-Hole Interactions in Anion Recognition. *Chem* **2018**, *4*, 731–783, doi:10.1016/j.chempr.2018.02.022.

5. Yu, B.; Pletka, C.C.; Iwahara, J. Quantifying and Visualizing Weak Interactions between Anions and Proteins. *Proceedings of the National Academy of Sciences* **2021**, *118*, e2015879118, doi:10.1073/pnas.2015879118.

6. Applications of Supramolecular Anion Recognition | Chemical Reviews Available online: https://pubs.acs.org/doi/10.1021/acs.chemrev.5b00099 (accessed on 18 October 2022).

7. Wu, X.; Gilchrist, A.M.; Gale, P.A. Prospects and Challenges in Anion Recognition and Transport. *Chem* **2020**, *6*, 1296–1309, doi:10.1016/j.chempr.2020.05.001.

8. Langton, M.J.; Serpell, C.J.; Beer, P.D. Anion Recognition in Water: Recent Advances from a Supramolecular and Macromolecular Perspective. *Angewandte Chemie International Edition* **2016**, *55*, 1974–1987, doi:10.1002/anie.201506589.

9. Groom, C.R.; Bruno, I.J.; Lightfoot, M.P.; Ward, S.C. The Cambridge Structural Database. *Acta Cryst B* **2016**, *72*, 171–179, doi:10.1107/S2052520616003954.

10. Berger, R.; Duff, K.; Leighton, J.L. Enantioselective Allylation of Ketone-Derived Benzoylhydrazones: Practical Synthesis of Tertiary Carbinamines. *J. Am. Chem. Soc.* **2004**, *126*, 5686–5687, doi:10.1021/ja0486418.

11. Wieghardt, K.; Kleine-Boymann, M.; Nuber, B.; Weiss, J.; Zsolnai, L.; Huttner, G. Macrocyclic Complexes of Lead(II): Crystal Structures of LPb(ClO4)2 and LPb(NO3)2 (L = 1,4,7-Triazacyclononane). *Inorg. Chem.* **1986**, *25*, 1647–1650, doi:10.1021/ic00230a024.

12. Dai, X.; Choi, S.-B.; Braun, C.W.; Vaidya, P.; Kilina, S.; Ugrinov, A.; Schulz, D.L.; Boudjouk, P. Halide Coordination of Perhalocyclohexasilane Si6X12 (X = Cl or Br). *Inorg. Chem.* **2011**, *50*, 4047–4053, doi:10.1021/ic102535n.

13. Teichmann, J.; Köstler, B.; Tillmann, J.; Moxter, M.; Kupec, R.; Bolte, M.; Lerner, H.-W.; Wagner, M. Halide-Ion Diadducts of Perhalogenated Cyclopenta- and Cyclohexasilanes. *Zeitschrift für anorganische und allgemeine Chemie* **2018**, *644*, 956–962, doi:10.1002/zaac.201800145.

14. Tillmann, J.; Meyer, L.; Schweizer, J.I.; Bolte, M.; Lerner, H.-W.; Wagner, M.; Holthausen, M.C. Chloride-Induced Aufbau of Perchlorinated Cyclohexasilanes from Si2Cl6: A Mechanistic Scenario. *Chemistry – A European Journal* **2014**, *20*, 9234–9239, doi:10.1002/chem.201402655.

15. Bamberg, M.; Bursch, M.; Hansen, A.; Brandl, M.; Sentis, G.; Kunze, L.; Bolte, M.; Lerner, H.-W.; Grimme, S.; Wagner, M. [Cl@Si20H20]−: Parent Siladodecahedrane with Endohedral Chloride Ion. *J. Am. Chem. Soc.* **2021**, *143*, 10865–10871, doi:10.1021/jacs.1c05598.

16. Taylor, M.S. Anion Recognition Based on Halogen, Chalcogen, Pnictogen and Tetrel Bonding. *Coordination Chemistry Reviews* **2020**, *413*, 213270, doi:10.1016/j.ccr.2020.213270.

17. Scheiner, S. Tetrel Bonding as a Vehicle for Strong and Selective Anion Binding. *Molecules* **2018**, *23*, 1147, doi:10.3390/molecules23051147.

18. Scheiner, S.; Michalczyk, M.; Zierkiewicz, W. Coordination of Anions by Noncovalently Bonded σ-Hole Ligands. *Coordination Chemistry Reviews* **2020**, *405*, 213136, doi:10.1016/j.ccr.2019.213136.

19. Molina, P.; Zapata, F.; Caballero, A. Anion Recognition Strategies Based on Combined Noncovalent Interactions. *Chem. Rev.* **2017**, *117*, 9907–9972, doi:10.1021/acs.chemrev.6b00814.

20. Liu, Y.-Z.; Yuan, K.; Lv, L.-L.; Zhu, Y.-C.; Yuan, Z. Designation and Exploration of Halide–Anion Recognition Based on Cooperative Noncovalent Interactions Including Hydrogen Bonds and Anion−π. *J. Phys. Chem. A* **2015**, *119*, 5842–5852, doi:10.1021/acs.jpca.5b02952.

21. Esrafili, M.D.; Mousavian, P. Strong Tetrel Bonds: Theoretical Aspects and Experimental Evidence. *Molecules* **2018**, *23*, 2642, doi:10.3390/molecules23102642.

22. Bartashevich, E.; Matveychuk, Y.; Tsirelson, V. Identification of the Tetrel Bonds between Halide Anions and Carbon Atom of Methyl Groups Using Electronic Criterion. *Molecules* **2019**, *24*, 1083, doi:10.3390/molecules24061083.

23. Grabowski, S.J. Tetrel Bond–σ-Hole Bond as a Preliminary Stage of the SN2 Reaction. *Phys. Chem. Chem. Phys.* **2014**, *16*, 1824–1834, doi:10.1039/C3CP53369G.

24. Bartashevich, E.V.; Mukhitdinova, S.E.; Klyuev, I.V.; Tsirelson, V.G. Can We Merge the Weak and Strong Tetrel Bonds? Electronic Features of Tetrahedral Molecules Interacted with Halide Anions. *Molecules* **2022**, *27*, 5411, doi:10.3390/molecules27175411.

25. Scheiner, S. Origins and Properties of the Tetrel Bond. *Phys. Chem. Chem. Phys.* **2021**, *23*, 5702–5717, doi:10.1039/D1CP00242B.

26. Varadwaj, P.R.; Varadwaj, A.; Jin, B.-Y. Significant Evidence of C⋯O and C⋯C Long-Range Contacts in Several Heterodimeric Complexes of CO with CH3–X, Should One Refer to Them as Carbon and Dicarbon Bonds! *Phys. Chem. Chem. Phys.* **2014**, *16*, 17238–17252, doi:10.1039/C4CP01775G.

27. Mani, D.; Arunan, E. The X–C⋯Y (X = O/F, Y = O/S/F/Cl/Br/N/P) 'Carbon Bond' and Hydrophobic Interactions. *Phys. Chem. Chem. Phys.* **2013**, *15*, 14377–14383, doi:10.1039/C3CP51658J.

28. Matczak, P. Theoretical Investigation of the N → Sn Coordination in (Me3SnCN)2. *Struct Chem* **2015**, *26*, 301–318, doi:10.1007/s11224-014-0485-4.

29. Varadwaj, P.R.; Varadwaj, A.; Marques, H.M.; Yamashita, K. Definition of the Tetrel Bond. Https://Arxiv.Org/Abs/2210.10649 (Assessed on November 17, 2022) 2022.

30. Breneman, C.M.; Martinov, M. 3 - The Use of Electrostatic Potential Fields in QSAR and QSPR. In *Theoretical and Computational Chemistry*; Murray, J.S., Sen, K., Eds.; Molecular Electrostatic Potentials; Elsevier, 1996; Vol. 3, pp. 143–179.

31. Murray, J.S.; Politzer, P. The Molecular Electrostatic Potential: A Tool for Understanding and Predicting Molecular Interactions. In *Molecular Orbital Calculations for Biological Systems*; Sapse, A.-M., Ed.; Oxford University Press, 1998; p. 0 ISBN 978-0-19-509873-0.

32. Politzer, P.; Murray, J.S. Molecular Electrostatic Potentials: Significance and Applications. In *Chemical Reactivity in Confined Systems*; John Wiley & Sons, Ltd, 2021; pp. 113–134 ISBN 978-1-119-68335-3.

33. Politzer, P.; Murray, J.S. Quantitative Analyses of Molecular Surface Electrostatic Potentials in Relation to Hydrogen Bonding and Co-Crystallization. *Crystal Growth & Design* **2015**, *15*, 3767–3774, doi:10.1021/acs.cgd.5b00419.

34. Varadwaj, A.; Marques, H.M.; Varadwaj, P.R. Nature of Halogen-Centered Intermolecular Interactions in Crystal Growth and Design: Fluorine-Centered Interactions in Dimers in Crystalline Hexafluoropropylene as a Prototype. *Journal of Computational Chemistry* **2019**, *40*, 1836–1860, doi:10.1002/jcc.25836.

35. Varadwaj, A.; Varadwaj, P.R.; Marques, H.M.; Yamashita, K. The Pnictogen Bond, Together with Other Non-Covalent Interactions, in the Rational Design of One-, Two- and Three-Dimensional Organic-Inorganic Hybrid

Metal Halide Perovskite Semiconducting Materials, and Beyond. *International Journal of Molecular Sciences* **2022**, *23*, 8816, doi:10.3390/ijms23158816.

36. Varadwaj, P.R.; Varadwaj, A.; Marques, H.M.; Yamashita, K. The Phosphorus Bond, or the Phosphorus-Centered Pnictogen Bond: The Covalently Bound Phosphorus Atom in Molecular Entities and Crystals as a Pnictogen Bond Donor. *Molecules* **2022**, *27*, 1487, doi:10.3390/molecules27051487.
37. Varadwaj, P.R.; Varadwaj, A.; Marques, H.M. Halogen Bonding: A Halogen-Centered Noncovalent Interaction Yet to Be Understood. *Inorganics* **2019**, *7*, 40, doi:10.3390/inorganics7030040.
38. Varadwaj, P.R.; Varadwaj, A.; Marques, H.M. Does Chlorine in CH3Cl Behave as a Genuine Halogen Bond Donor? *Crystals* **2020**, *10*, 146, doi:10.3390/cryst10030146.
39. Varadwaj, P.R.; Varadwaj, A.; Marques, H.M.; Yamashita, K. The Pnictogen Bond Formation Ability of Bonded Bismuth Atoms in Molecular Entities in the Crystalline Phase: A Perspective. Https://Arxiv.Org/Abs/2209.07319 (Assessed on November, 17, 2022). 2022.
40. Bader, R.F.W. Atoms in Molecules Available online: https://pubs.acs.org/doi/pdf/10.1021/ar00109a003 (accessed on 19 October 2022).
41. Bader, R.F.W. Atoms in Molecules. In *Encyclopedia of Computational Chemistry*; John Wiley & Sons, Ltd, 2002 ISBN 978-0-470-84501-1.
42. Bader, R.F.W.; Bayles, D. Properties of Atoms in Molecules:  Group Additivity. *J. Phys. Chem. A* **2000**, *104*, 5579–5589, doi:10.1021/jp9943631.
43. Bader, R.F.W.; Nguyen-Dang, T.T. Quantum Theory of Atoms in Molecules–Dalton Revisited. In *Advances in Quantum Chemistry*; Löwdin, P.-O., Ed.; Academic Press, 1981; Vol. 14, pp. 63–124.
44. Lefebvre, C.; Khartabil, H.; Boisson, J.-C.; Contreras-García, J.; Piquemal, J.-P.; Hénon, E. The Independent Gradient Model: A New Approach for Probing Strong and Weak Interactions in Molecules from Wave Function Calculations. *ChemPhysChem* **2018**, *19*, 724–735, doi:10.1002/cphc.201701325.
45. Lefebvre, C.; Khartabil, H.; Boisson, J.-C.; Contreras-García, J.; Piquemal, J.-P.; Hénon, E. The Independent Gradient Model: A New Approach for Probing Strong and Weak Interactions in Molecules from Wave Function Calculations. *Chemphyschem* **2018**, *19*, 724–735, doi:10.1002/cphc.201701325.
46. Varadwaj, P.R.; Varadwaj, A.; Marques, H.M.; Yamashita, K. The Nitrogen Bond, or the Nitrogen-Centered Pnictogen Bond: The Covalently Bound Nitrogen Atom in Molecular Entities and Crystals as a Pnictogen Bond Donor. *Compounds* **2022**, *2*, 80–110, doi:10.3390/compounds2010007.
47. Chai, J.-D.; Head-Gordon, M. Long-Range Corrected Hybrid Density Functionals with Damped Atom–Atom Dispersion Corrections. *Phys. Chem. Chem. Phys.* **2008**, *10*, 6615–6620, doi:10.1039/B810189B.
48. Frisch, M.J.; Head-Gordon, M.; Pople, J.A. A Direct MP2 Gradient Method. *Chemical Physics Letters* **1990**, *166*, 275–280, doi:10.1016/0009-2614(90)80029-D.
49. Frisch, M.J.; Head-Gordon, M.; Pople, J.A. Semi-Direct Algorithms for the MP2 Energy and Gradient. *Chemical Physics Letters* **1990**, *166*, 281–289, doi:10.1016/0009-2614(90)80030-H.
50. Head-Gordon, M.; Head-Gordon, T. Analytic MP2 Frequencies without Fifth-Order Storage. Theory and Application to Bifurcated Hydrogen Bonds in the Water Hexamer. *Chemical Physics Letters* **1994**, *220*, 122–128, doi:10.1016/0009-2614(94)00116-2.
51. Purvis, G.D.; Bartlett, R.J. A Full Coupled-cluster Singles and Doubles Model: The Inclusion of Disconnected Triples. *J. Chem. Phys.* **1982**, *76*, 1910–1918, doi:10.1063/1.443164.
52. Scuseria, G.E.; Janssen, C.L.; Schaefer, H.F., III An Efficient Reformulation of the Closed-Shell Coupled Cluster Single and Double Excitation (CCSD) Equations. *Journal of Chemical Physics* **1988**, *89*, 7382–7387, doi:10.1063/1.455269.

53. Gaussian 16, Revision C.01, Frisch, M. J.; Trucks, G. W.; Schlegel, H. B.; Scuseria, G. E.; Robb, M. A.; Cheeseman, J. R.; Scalmani, G.; Barone, V.; Petersson, G. A.; Nakatsuji, H.; Li, X.; Caricato, M.; Marenich, A. V.; Bloino, J.; Janesko, B. G.; Gomperts, R.; Mennucci, B.; Hratchian, H. P.; Ortiz, J. V.; Izmaylov, A. F.; Sonnenberg, J. L.; Williams-Young, D.; Ding, F.; Lipparini, F.; Egidi, F.; Goings, J.; Peng, B.; Petrone, A.; Henderson, T.; Ranasinghe, D.; Zakrzewski, V. G.; Gao, J.; Rega, N.; Zheng, G.; Liang, W.; Hada, M.; Ehara, M.; Toyota, K.; Fukuda, R.; Hasegawa, J.; Ishida, M.; Nakajima, T.; Honda, Y.; Kitao, O.; Nakai, H.; Vreven, T.; Throssell, K.; Montgomery, J. A., Jr.; Peralta, J. E.; Ogliaro, F.; Bearpark, M. J.; Heyd, J. J.; Brothers, E. N.; Kudin, K. N.; Staroverov, V. N.; Keith, T. A.; Kobayashi, R.; Normand, J.; Raghavachari, K.; Rendell, A. P.; Burant, J. C.; Iyengar, S. S.; Tomasi, J.; Cossi, M.; Millam, J. M.; Klene, M.; Adamo, C.; Cammi, R.; Ochterski, J. W.; Martin, R. L.; Morokuma, K.; Farkas, O.; Foresman, J. B.; Fox, D. J. Gaussian, Inc., Wallingford CT, 2016. Available online: https://gaussian.com/citation/ (accessed on 19 October 2022).

54. Schuchardt, K.L.; Didier, B.T.; Elsethagen, T.; Sun, L.; Gurumoorthi, V.; Chase, J.; Li, J.; Windus, T.L. Basis Set Exchange: A Community Database for Computational Sciences. *J. Chem. Inf. Model.* **2007**, *47*, 1045–1052, doi:10.1021/ci600510j.

55. Pritchard, B.P.; Altarawy, D.; Didier, B.; Gibson, T.D.; Windus, T.L. New Basis Set Exchange: An Open, Up-to-Date Resource for the Molecular Sciences Community. *J. Chem. Inf. Model.* **2019**, *59*, 4814–4820, doi:10.1021/acs.jcim.9b00725.

56. Wu, Q.; Xie, X.; Li, Q.; Scheiner, S. Enhancement of Tetrel Bond Involving Tetrazole-TtR3 (Tt = C, Si; R = H, F). Promotion of SiR3 Transfer by a Triel Bond. *Phys. Chem. Chem. Phys.* **2022**, doi:10.1039/D2CP04194D.

57. An, X.; Yang, X.; Li, Q. Tetrel Bonds between Phenyltrifluorosilane and Dimethyl Sulfoxide: Influence of Basis Sets, Substitution and Competition. *Molecules* **2021**, *26*, 7231, doi:10.3390/molecules26237231.

58. Varadwaj, A.; Marques, H.M.; Varadwaj, P.R. Is the Fluorine in Molecules Dispersive? Is Molecular Electrostatic Potential a Valid Property to Explore Fluorine-Centered Non-Covalent Interactions? *Molecules* **2019**, *24*, 379, doi:10.3390/molecules24030379.

59. Bader, R.F.W.; Carroll, M.T.; Cheeseman, J.R.; Chang, C. Properties of Atoms in Molecules: Atomic Volumes. *J. Am. Chem. Soc.* **1987**, *109*, 7968–7979, doi:10.1021/ja00260a006.

60. Varadwaj, P.R.; Varadwaj, A.; Marques, H.M.; Yamashita, K. Can Combined Electrostatic and Polarization Effects Alone Explain the F···F Negative-Negative Bonding in Simple Fluoro-Substituted Benzene Derivatives? A First-Principles Perspective. *Computation* **2018**, *6*, 51, doi:10.3390/computation6040051.

61. Varadwaj, P.R. Does Oxygen Feature Chalcogen Bonding? *Molecules* **2019**, *24*, 3166, doi:10.3390/molecules24173166.

62. Varadwaj, A.; Varadwaj, P.R.; Marques, H.M.; Yamashita, K. The Stibium Bond or the Antimony-Centered Pnictogen Bond: The Covalently Bound Antimony Atom in Molecular Entities in Crystal Lattices as a Pnictogen Bond Donor. *International Journal of Molecular Sciences* **2022**, *23*, 4674, doi:10.3390/ijms23094674.

63. Lu, T.; Chen, F. Multiwfn: A Multifunctional Wavefunction Analyzer. *Journal of Computational Chemistry* **2012**, *33*, 580–592, doi:10.1002/jcc.22885.

64. Humphrey, W.; Dalke, A.; Schulten, K. VMD: Visual Molecular Dynamics. *Journal of Molecular Graphics* **1996**, *14*, 33–38, doi:10.1016/0263-7855(96)00018-5.

65. Politzer, P.; Murray, J.S. σ-Hole Interactions: Perspectives and Misconceptions. *Crystals* **2017**, *7*, 212, doi:10.3390/cryst7070212.

66. Politzer, P.; Murray, J.S.; Clark, T.; Resnati, G. The σ-Hole Revisited. *Phys. Chem. Chem. Phys.* **2017**, *19*, 32166–32178, doi:10.1039/C7CP06793C.

67. σ-Hole Bonding: Molecules Containing Group VI Atoms | SpringerLink Available online: https://link.springer.com/article/10.1007/s00894-007-0225-4 (accessed on 9 November 2022).

68. Clark, T.; Hennemann, M.; Murray, J.S.; Politzer, P. Halogen Bonding: The Sigma-Hole. Proceedings of "Modeling Interactions in Biomolecules II", Prague, September 5th-9th, 2005. *J Mol Model* **2007**, *13*, 291–296, doi:10.1007/s00894-006-0130-2.
69. Boys, S.F.; Bernardi, F. The Calculation of Small Molecular Interactions by the Differences of Separate Total Energies. Some Procedures with Reduced Errors. *Molecular Physics* **1970**, *19*, 553–566, doi:10.1080/00268977000101561.
70. AIMAll (Version 19.10.12), Todd A. Keith, TK Gristmill Software, Overland Park KS, USA, 2019 (Aim.Tkgristmill.Com) Available online: http://aim.tkgristmill.com/references.html (accessed on 24 October 2022).
71. Lefebvre, C.; Rubez, G.; Khartabil, H.; Boisson, J.-C.; Contreras-García, J.; Hénon, E. Accurately Extracting the Signature of Intermolecular Interactions Present in the NCI Plot of the Reduced Density Gradient versus Electron Density. *Phys. Chem. Chem. Phys.* **2017**, *19*, 17928–17936, doi:10.1039/C7CP02110K.
72. The Independent Gradient Model: A New Approach for Probing Strong and Weak Interactions in Molecules from Wave Function Calculations - Lefebvre - 2018 - ChemPhysChem - Wiley Online Library Available online: https://chemistry-europe.onlinelibrary.wiley.com/doi/10.1002/cphc.201701325 (accessed on 19 October 2022).
73. Multiwfn Available online: http://sobereva.com/multiwfn/ (accessed on 24 October 2022).
74. Fradera, X.; Austen, M.A.; Bader, R.F.W. The Lewis Model and Beyond. *J. Phys. Chem. A* **1999**, *103*, 304–314, doi:10.1021/jp983362q.
75. Outeiral, C.; Vincent, M.A.; Martín Pendás, Á.; Popelier, P.L.A. Revitalizing the Concept of Bond Order through Delocalization Measures in Real Space †Electronic Supplementary Information (ESI) Available. See DOI: 10.1039/C8sc01338a. *Chem Sci* **2018**, *9*, 5517–5529, doi:10.1039/c8sc01338a.
76. Matito, E.; Poater, J.; Solà, M.; Duran, M.; Salvador, P. Comparison of the AIM Delocalization Index and the Mayer and Fuzzy Atom Bond Orders. *J. Phys. Chem. A* **2005**, *109*, 9904–9910, doi:10.1021/jp0538464.
77. Firme, C.L.; Antunes, O.A.C.; Esteves, P.M. Relation between Bond Order and Delocalization Index of QTAIM. *Chemical Physics Letters* **2009**, *468*, 129–133, doi:10.1016/j.cplett.2008.12.004.
78. Kaduk, J.A. Use of the Inorganic Crystal Structure Database as a Problem Solving Tool. *Acta Cryst B* **2002**, *58*, 370–379, doi:10.1107/S0108768102003476.
79. Hellenbrandt, M. The Inorganic Crystal Structure Database (ICSD)—Present and Future. *Crystallography Reviews* **2004**, *10*, 17–22, doi:10.1080/08893110410001664882.
80. Belsky, A.; Hellenbrandt, M.; Karen, V.L.; Luksch, P. New Developments in the Inorganic Crystal Structure Database (ICSD): Accessibility in Support of Materials Research and Design. *Acta Cryst B* **2002**, *58*, 364–369, doi:10.1107/S0108768102006948.
81. Allmann, R.; Hinek, R. The Introduction of Structure Types into the Inorganic Crystal Structure Database ICSD. *Acta Cryst A* **2007**, *63*, 412–417, doi:10.1107/S0108767307038081.
82. Inorganic Crystal Structure Database Available online: https://icsd.products.fiz-karlsruhe.de/en (accessed on 21 October 2022).
83. Cremer, D.; Kraka, E. A Description of the Chemical Bond in Terms of Local Properties of Electron Density and Energy. *Croatica Chemica Acta* **1984**, *57*, 1259–1281.
84. Grabowski, S.J. The Nature of Triel Bonds, a Case of B and Al Centres Bonded with Electron Rich Sites. *Molecules* **2020**, *25*, 2703, doi:10.3390/molecules25112703.
85. Inorganics | Free Full-Text | Chalcogen⋯Chalcogen Bonding in Molybdenum Disulfide, Molybdenum Diselenide and Molybdenum Ditelluride Dimers as Prototypes for a Basic Understanding of the Local Interfacial Chemical Bonding Environment in 2D Layered Transition Metal Dichalcogenides | HTML Available online: https://www.mdpi.com/2304-6740/10/1/11/htm (accessed on 21 October 2022).

86. Varadwaj, P.; Varadwaj, A.; Marques, H. *Very Strong Chalcogen Bonding: Is Oxygen in Molecules Capable of Forming It? A First-Principles Perspective*; Preprints, 2020;
87. Varadwaj, P.R.; Varadwaj, A.; Marques, H.M. DFT-B3LYP, NPA-, and QTAIM-Based Study of the Physical Properties of [M(II)(H2O)2(15-Crown-5)] (M = Mn, Fe, Co, Ni, Cu, Zn) Complexes. *J. Phys. Chem. A* **2011**, *115*, 5592–5601, doi:10.1021/jp2001157.
88. Zhurova, E.A.; Tsirelson, V.G. Electron Density and Energy Density View on the Atomic Interactions in SrTiO3. *Acta Cryst B* **2002**, *58*, 567–575, doi:10.1107/S0108768102009692.
89. Schwerdtfeger, P.; Nagle, J.K. 2018 Table of Static Dipole Polarizabilities of the Neutral Elements in the Periodic Table. *Molecular Physics* **2019**, *117*, 1200–1225, doi:10.1080/00268976.2018.1535143.
90. Grabowski, S.J. Lewis Acid Properties of Tetrel Tetrafluorides—The Coincidence of the σ-Hole Concept with the QTAIM Approach. *Crystals* **2017**, *7*, 43, doi:10.3390/cryst7020043.
91. Jabłoński, M. Does the Presence of a Bond Path Really Mean Interatomic Stabilization? The Case of the Ng@Superphane (Ng = He, Ne, Ar, and Kr) Endohedral Complexes. *Symmetry* **2021**, *13*, 2241, doi:10.3390/sym13122241.
92. Lane, J.R.; Contreras-García, J.; Piquemal, J.-P.; Miller, B.J.; Kjaergaard, H.G. Are Bond Critical Points Really Critical for Hydrogen Bonding? *J. Chem. Theory Comput.* **2013**, *9*, 3263–3266, doi:10.1021/ct400420r.
93. Thomsen, D.L.; Axson, J.L.; Schrøder, S.D.; Lane, J.R.; Vaida, V.; Kjaergaard, H.G. Intramolecular Interactions in 2-Aminoethanol and 3-Aminopropanol. *J. Phys. Chem. A* **2013**, *117*, 10260–10273, doi:10.1021/jp405512y.
94. Arunan, E.; Desiraju, G.R.; Klein, R.A.; Sadlej, J.; Scheiner, S.; Alkorta, I.; Clary, D.C.; Crabtree, R.H.; Dannenberg, J.J.; Hobza, P.; et al. Definition of the Hydrogen Bond (IUPAC Recommendations 2011). *Pure and Applied Chemistry* **2011**, *83*, 1637–1641, doi:10.1351/PAC-REC-10-01-02.
95. Desiraju, G.R.; Ho, P.S.; Kloo, L.; Legon, A.C.; Marquardt, R.; Metrangolo, P.; Politzer, P.; Resnati, G.; Rissanen, K. Definition of the Halogen Bond (IUPAC Recommendations 2013). *Pure and Applied Chemistry* **2013**, *85*, 1711–1713, doi:10.1351/PAC-REC-12-05-10.
96. Aakeroy, C.B.; Bryce, D.L.; Desiraju, G.R.; Frontera, A.; Legon, A.C.; Nicotra, F.; Rissanen, K.; Scheiner, S.; Terraneo, G.; Metrangolo, P.; et al. Definition of the Chalcogen Bond (IUPAC Recommendations 2019). *Pure and Applied Chemistry* **2019**, *91*, 1889–1892, doi:10.1515/pac-2018-0713.
97. Varadwaj, A.; Varadwaj, P.R.; Marques, H.M.; Yamashita, K. Definition of the Pnictogen Bond: A Perspective. *Inorganics* **2022**, *10*, 149, doi:10.3390/inorganics10100149.
98. Liu, N.; Li, Q.; Scheiner, S.; Xie, X. Resonance-Assisted Intramolecular Triel Bonds. *Phys. Chem. Chem. Phys.* **2022**, *24*, 15015–15024, doi:10.1039/D2CP01244H.
99. Alvarez, S. A Cartography of the van Der Waals Territories. *Dalton Trans.* **2013**, *42*, 8617–8636, doi:10.1039/C3DT50599E.
100. Astatine | At (Element) - PubChem Available online: https://pubchem.ncbi.nlm.nih.gov/element/Astatine#section=Atomic-Weight (accessed on 25 October 2022).
101. Politzer, P.; Murray, J.S. The Use and Misuse of van Der Waals Radii. *Struct Chem* **2021**, *32*, 623–629, doi:10.1007/s11224-020-01713-7.
102. Varadwaj, P.R.; Varadwaj, A.; Jin, B.-Y. Hexahalogenated and Their Mixed Benzene Derivatives as Prototypes for the Understanding of Halogen···halogen Intramolecular Interactions: New Insights from Combined DFT, QTAIM-, and RDG-Based NCI Analyses. *Journal of Computational Chemistry* **2015**, *36*, 2328–2343, doi:10.1002/jcc.24211.
103. Johansson, M.P.; Swart, M. Intramolecular Halogen–Halogen Bonds? *Phys. Chem. Chem. Phys.* **2013**, *15*, 11543–11553, doi:10.1039/C3CP50962A.

104. Xu, H.; Cheng, J.; Yang, X.; Liu, Z.; Li, W.; Li, Q. Comparison of σ-Hole and π-Hole Tetrel Bonds Formed by Pyrazine and 1,4-Dicyanobenzene: The Interplay between Anion–π and Tetrel Bonds. *ChemPhysChem* **2017**, *18*, 2442–2450, doi:10.1002/cphc.201700660.
105. Van der Maelen, J.F.; Cabeza, J.A. A Topological Analysis of the Bonding in [M2(CO)10] and [M3(μ-H)3(CO)12] Complexes (M = Mn, Tc, Re). *Theor Chem Acc* **2016**, *135*, 64, doi:10.1007/s00214-016-1821-0.
106. Chapter 1 Hydrogen Bond – Definitions, Criteria of Existence and Various Types. **2020**, 1–40, doi:10.1039/9781839160400-00001.
107. Desiraju, G.R. Hydrogen Bridges in Crystal Engineering: Interactions without Borders. *Acc. Chem. Res.* **2002**, *35*, 565–573, doi:10.1021/ar010054t.
108. Varadwaj, A.; Varadwaj, P.R.; Yamashita, K. Hybrid Organic–Inorganic CH3NH3PbI3 Perovskite Building Blocks: Revealing Ultra-Strong Hydrogen Bonding and Mulliken Inner Complexes and Their Implications in Materials Design. *Journal of Computational Chemistry* **2017**, *38*, 2802–2818, doi:10.1002/jcc.25073.